%% file: wires9.tex
\documentclass[pra,twocolumn,letterpaper,floatfix]{revtex4}

\usepackage{amssymb,amsmath,amsthm,amsfonts}
\usepackage{graphicx}
\usepackage{epsfig}
\usepackage{subfigure}
\usepackage{comment}
\usepackage{tikz}
\usetikzlibrary{backgrounds,fit,decorations.pathreplacing,decorations.pathmorphing,positioning,calc,fadings}  % TikZ libraries

\newcommand{\field}[1]{\mathbb{#1}}

\newcommand{\R}{\field{R}}
\newcommand{\Z}{\field{Z}}

\newcommand{\ket}[1]{\ensuremath{| #1 \rangle}}
\newcommand{\bra}[1]{\ensuremath{\langle #1 |}}
\newcommand{\ketbra}[2]{\ensuremath{| #1 \rangle \langle #2 |}}
\newcommand{\braket}[2]{\ensuremath{\langle #1 | #2 \rangle}}
\newcommand{\set}[1]{\ensuremath{\left \{ #1 \right \}}}

\newcommand{\mtxtwotwo}[4]{\ensuremath{\left[ \begin{array}{cc} #1 & #2 \\ #3 & #4 \end{array} \right]}}

\newcommand{\brackets}[1]{\ensuremath{\left( #1 \right)}}
\newcommand{\sbrackets}[1]{\ensuremath{\left[ #1 \right]}}
\newcommand{\abs}[1]{\ensuremath{\left| #1 \right|}}

\newcommand{\gl}{\ensuremath{\mathrm{GL}\brackets{2,\mathbb{C}}}}
\newcommand{\rx}[1]{\ensuremath{\mathrm{R}_x\sbrackets{#1}}}

\newcommand{\rz}[1]{\ensuremath{\mathrm{R}_z\sbrackets{#1}}}
\newcommand{\rxs}[1]{\ensuremath{\mathrm{R}_x\sbrackets{#1}}}

\newcommand{\rzs}[1]{\ensuremath{\mathrm{R}_z\sbrackets{#1}}}
\newcommand{\hrz}[1]{\ensuremath{\mathrm{H}\rz{#1}}}
\newcommand{\cl}[1]{\ensuremath{\ket{\textrm{Cl}_{#1}}}}
\newcommand{\cz}[2]{\ensuremath{\mathrm{CZ}}}

\newcommand{\X}{\ensuremath{\mathrm{X}}}
\newcommand{\Y}{\ensuremath{\mathrm{Y}}}
\renewcommand{\Z}{\ensuremath{\mathrm{Z}}}
\renewcommand{\H}{\ensuremath{\mathrm{H}}}
\newcommand{\N}{\ensuremath{\mathrm{N}}}
\newcommand{\B}{\ensuremath{\mathrm{B}}}

\newcommand{\U}{\ensuremath{\mathrm{U}}}
\newcommand{\V}{\ensuremath{\mathrm{V}}}
\newcommand{\M}{\ensuremath{\mathrm{M}}}
\newcommand{\Mp}{\ensuremath{\mathrm{M^{\perp}}}}
\renewcommand{\R}{\ensuremath{\mathrm{R}}}
\newcommand{\I}{\ensuremath{\mathrm{I}}}
\newcommand{\p}{\ensuremath{\prime}}

\newtheorem{thm}{Theorem}[section]

\newtheorem{lem}[thm]{Lemma}
\newtheorem{defn}[thm]{Definition}

\theoremstyle{remark}

\theoremstyle{definition}

\begin{document} 

%\title{Strategies for Measurement-Based Quantum Computation with SLOCC-Transformed Cluster States}

\title{Strategies for measurement-based quantum computation with cluster states
transformed by stochastic local operations and classical communication}

\author{Adam~G.~D'Souza and David~L.~Feder} 
\affiliation{Department of Physics and Astronomy and Institute for Quantum
Information Science, University of Calgary, Calgary, Alberta, Canada}

\date{\today}

\begin{abstract}
%Universal quantum computation can be accomplished via projective single-qubit measurements on a highly entangled resource state, together with classical feedforward of the measurement results. The best-known example of such a resource state is the cluster state, on which judiciously chosen single-qubit measurements can be used to simulate an arbitrary quantum circuit with a number of measurements that is linear in the number of gates. 
We examine cluster states transformed by stochastic local operations and classical communication, as a resource for deterministic universal computation driven strictly by projective measurements. We identify circumstances under which such states in one dimension constitute resources for random-length single-qubit rotations, in one case quasi-deterministically ($\N-\U-\N$ states) and in another probabilistically ($\B-\U-\B$ states). In contrast to the cluster states, the $\N-\U-\N$ states exhibit spin correlation functions that decay exponentially with distance, while the $\B-\U-\B$ states can be arbitrarily locally pure. A two-dimensional square $\N-\U-\N$ lattice is a universal resource for quasi-deterministic measurement-based quantum computation. Measurements on cubic $\B-\U-\B$ states yield two-dimensional cluster states with bond defects, whose connectivity exceeds the percolation threshold for a critical value of the local purity.
\end{abstract}

\maketitle

\section{Introduction}
\label{sec:introduction}
In the Measurement-Based Quantum Computation (MBQC) 
model~\cite{Raussendorf2001,Raussendorf2003}, one starts with a highly entangled many-qubit quantum state called a resource state, and processes logical information via single-qubit measurements on the physical qubits of the resource state. In order to compensate for the randomness of the measurement outcomes, the bases in which measurements are performed must be conditioned on the outcomes of previous measurements. Proceeding in this way, one can teleport logical quantum information situated on one part of the state to another part, but having been subjected to some desired unitary transformation. If the basic unitary transformations that can be applied via single-qubit measurements on the resource state generate a set that is dense in SU(2), then the resource is said to be universal (in the terminology of Ref.~\cite{VandenNest2007}, this is the notion of CQ-universality, and such a resource state would be called a universal state preparator).

The archetypal family of resource states known to be universal for efficient MBQC is the so-called cluster state~\cite{Raussendorf2001,Raussendorf2003}. This
state is special in several ways: all spin correlation functions are strictly nearest-neighbor~\cite{Gross2007, Gross2007a}, the localizable entanglement between any pair of qubits is maximal~\cite{Gross2007,Gross2007a}, it is the only state (up to local unitaries) on small system sizes that saturates the Tsallis and Renyi entropies of entanglement~\cite{Gour2010}, it cannot be the non-degenerate ground state of a two-body spin Hamiltonian~\cite{Nielsen2006,VandenNest2008}, and so on. One might expect that one or more of these properties would be necessarily satisfied by any universal resource state. This has turned out not to be the case; several authors in recent years have identified resources that differ materially from the cluster states~\cite{Bartlett2006,Gross2007,Gross2007a,Brennen2008,Chen2009,Doherty2009,Miyake2010,Wei2011}.

The newly discovered richness in the landscape of resources notwithstanding, the property of universality is exceedingly rare; not only must a family of universal resource states saturate various measures of entanglement in the thermodynamic limit~\cite{VandenNest2007,Mora2010}, but the entanglement with respect to other measures must not be too high~\cite{Gross2009,Bremner2009}. Therefore, it is highly unlikely that a random pure state will be universal, so a search for new resources must be heavily constrained in order to have a reasonable chance of success.

Recently, a number of new resources~\cite{Gross2007,Gross2007a,Brennen2008,Gross2010} have been proven to be universal by means of reduction to a known resource state~\cite{Chen2010}. The reduction strategies of interest are those composed purely of local operations, possibly augmented by classical communication. They are typically stochastic, in the sense that the known resource state is smaller than the original state. In other words, these resources all appear to be within the equivalence class of the cluster states under Stochastic Local Operations and Classical Communication (SLOCC). The SLOCC-equivalence class of an $n$-qubit pure state is known to be its orbit under $\gl^{\otimes n}$, the group of $n$-fold tensor products of two-by-two invertible matrices over the complex numbers. In other words, two $n$-qubit states \ket{\psi} and \ket{\phi} are equivalent under SLOCC if and only if
\begin{equation}
\ket{\psi} = S^{(1)} \otimes S_2 \otimes \dots \otimes S_n \ket{\phi}
\end{equation}
where the $\set{S_i}\in\gl$ are invertible, two-by-two complex matrices.

A natural question thus arises: are all universal MBQC resources SLOCC-equivalent to the cluster states? More precisely, is any $n$-qubit element of a family of universal resource states SLOCC-equivalent to an $n$-qubit cluster state?

In this paper, we tackle a related question, namely: what states in the SLOCC-equivalence class of the two-dimensional cluster states are universal for MBQC? It is clear, by construction, that each state in this class can be stochastically reduced to a cluster state, but what is not clear is whether it is possible to compute directly on the image of a cluster state under some invertible, local map. We show that there is a restricted subclass of invertible local transformations, strictly including the local unitaries, whose image is a set of quasi-deterministic resources for MBQC, where in general the computation is of random length and `repeat-until-success' strategies must be employed (c.f. Refs.~\cite{Gross2007a,Gross2010}). In particular, we identify two types of SLOCC operators whose action can in certain cases preserve the usefulness of the cluster state as a resource. The first type, which we call $\N$-type operators, comprises those operators that preserve the relative norms of the computational basis states. The second type, called $\B$-type operators, are those that preserve their orthogonality
(i.e.\ are in a sense basis-preserving).

In particular, we show that when $\N$-type operators act on alternating qubits in a 1D cluster state, the state remains a quasi-deterministic resource for single-qubit rotations. We refer to such 1D states as $\N-\U-\N$ chains. In contrast to the cluster state, the number of measurements required to implement an arbitrary single-qubit rotation with an $\N-\U-\N$ chain is random rather than fixed. Furthermore, the state exhibits non-zero spin-spin correlations that decay exponentially with distance. These properties are shared by other resources previously appearing in the literature~\cite{Gross2007,Brennen2008,Bartlett2010,Bartlett2010}, notably those based on the so-called AKLT model~\cite{Affleck1987,Affleck1988}. We also how 1D $\N-\U-\N$ chains can be coupled together to produce a quasi-deterministic 2D resource for universal MBQC.

Next, we show that when $\B$-type operators act on alternating qubits in a 1D cluster state, the result is in general a probabilistic resource for single-qubit rotations. We call these states $\B-\U-\B$ chains. We find that under a restricted subset of $\B$-type operators, the three-dimensional analogs of $\B-\U-\B$ chains constitute quasi-deterministic resources for MBQC under strictly projective measurements. A similar result was exhibited in Ref.~\cite{Mora2010}, in which a 2D cluster state deformed by $\B$-type operators was shown to be reducible to a percolated 2D cluster state~\cite{Kieling2007b} by the action of three-element POVMs. The $\B-\U-\B$ states have the interesting property that each qubit can be arbitrarily locally pure, or alternatively that an individual qubit can be arbitrarily weakly entangled with the rest of the state, as measured by the von Neumann entropy of entanglement. Like the cluster states, they also exhibit vanishing long-range correlations, with no spin-spin correlation functions beyond second-nearest-neighbor surviving.

The structure of the paper is as follows: in Section~\ref{sec:background}, we briefly review the theory of measurement-based quantum computation using cluster
states, and introduce the various definitions and notation used in the 
technical part of this paper. In Section~\ref{sec:results}, we describe the effects of invertible local operators acting on a cluster state on the class of linear transformations that can be logically implemented via adaptive single-qubit measurements on this new state, and outline some strategies for dealing with these effects. In Section~\ref{sec:randomlength}, we provide explicit examples of some structures of SLOCC-transformed cluster states that are universal for either probabilistic or deterministic single-qubit rotations or full MBQC. Finally, in Section~\ref{sec:discussion}, we discuss the relationship of our resource states with previously known quasi-deterministic resources, and outline the prospects of identifying hitherto unknown resource states by this method.

\section{Background}
\label{sec:background}
An $n$-qubit cluster state can be defined in terms of the stabilizer formalism~\cite{Gottesman1997} as the unique $n$-qubit pure state \cl{n} satisfying the $n$ conditions
\begin{equation}
\label{eq:stabconds} \X_i \bigotimes_{j \in \mathcal{N}(i)} \Z_j \cl{n} = \cl{n},
\end{equation}
where $i \in \set{1,\dots,n}$ labels a qubit, $\mathcal{N}(i)$ denotes the spatial neighbourhood of qubit $i$, and $\X_i$ and $\Z_i$ denote the standard single-qubit Pauli operators, given in the computational basis by
\begin{eqnarray*}
\X & = & \mtxtwotwo{0}{1}{1}{0};\\
\Z & = & \mtxtwotwo{1}{0}{0}{-1},
\end{eqnarray*}
acting on qubit $i$. Alternatively, the cluster state can be identified as the result of a dynamical process in which
\begin{enumerate}
\item $n$ qubits are initialized in the state $\ket{+}^{\otimes n}$, where $\ket{+} \equiv \frac{1}{\sqrt{2}} \brackets{\ket{0}+\ket{1}}$ is the $+1$-eigenstate of $\X$;
\item $\mathrm{CZ}$ entangling gates, whose action on the computational basis states is given by 
\begin{equation}
\mathrm{CZ}_{i,j}\ket{x,y}_{i,j} = \brackets{-1}^{x \cdot y} \ket{x,y}_{i,j},
\end{equation}
are applied between each pair of neighbouring qubits $(i,j)$. Thus,
\begin{equation}
\cl{n} = \prod_{\langle i,j \rangle} \mathrm{CZ}_{i,j} \ket{+}^{\otimes n},
\label{eq:defcluster}
\end{equation}
where $\langle i,j \rangle$ indicates that $i$ and $j$ label neighbouring qubits. 
\end{enumerate}

For notational convenience, define a global entangling operation on a lattice,
\begin{equation}
\mathfrak{G}_{k,l} := \displaystyle \prod_{j=k}^{l-1}\mathrm{CZ}_{j,j+1},
\end{equation}
to be the tensor product of $\mathrm{CZ}$ gates acting between all nearest-neighbour pairs of vertices on a line with labels between $k$ and $l$. Now consider a modified one-dimensional $n$-qubit cluster state,
\begin{equation}
	\label{eq:clusterdef}\ket{\textrm{Cl}_n^{1D}}' = \mathfrak{G}_{1,n} \ket{\psi}_1\ket{+}^{\otimes n-1}_{2,\dots,n}.
\end{equation}
where the first qubit was encoded in some general pure state \ket{\psi} before the global entangling operation $\mathfrak{G}_{1,n}$ was performed. The effect of projectively measuring the first qubit, the one on which \ket{\psi} was initially encoded, is to teleport the quantum information corresponding to the state \ket{\psi} to the next qubit, subject to some linear transformation depending upon the basis and outcome of the measurement. To see this, assume that $\ket{\psi} = a\ket{0} + b\ket{1}$. We then find that
\begin{eqnarray}
\nonumber \ket{\textrm{Cl}_n^{1D}}' & = & \mathfrak{G}_{2,n} \mathrm{CZ}_{1,2} \brackets{a\ket{0+}_{1,2} + b\ket{1+}_{1,2}} \ket{+}^{\otimes n-2}_{3,\dots,n} \\
\nonumber	& = & \mathfrak{G}_{2,n} \brackets{a\ket{0+}_{1,2} + b\ket{1-}_{1,2}} \ket{+}^{\otimes n-2}_{3,\dots,n} \\
\label{eq:clusteridentity}	& = & \mathfrak{G}_{2,n} \brackets{a\ket{0}_1 \I_2 + b\ket{1}_1 \Z_2} \ket{+}^{\otimes n-1}_{2,\dots,n}.
\end{eqnarray}

Projecting the first qubit via an arbitrary rank-1 projector \ketbra{m}{m}, the state of the system (neglecting the projected qubit and overall normalization) becomes

\begin{eqnarray*}
& \ket{\Phi} & =  \mathfrak{G}_{2,n} \brackets{a\braket{m}{0} \I_2 + b\braket{m}{1} \Z_2} \ket{+}^{\otimes n-1}_{2,\dots,n} \\
			& = & \mathfrak{G}_{2,n} \brackets{a\braket{m}{0} \ket{+}_2 + b\braket{m}{1} \ket{-}_2} \ket{+}^{\otimes n-2}_{3,\dots,n} \\
			& = & \mathfrak{G}_{2,n} \H_2 \brackets{a\braket{m}{0} \ket{0}_2 + b\braket{m}{1} \ket{1}_2} \ket{+}^{\otimes n-2}_{3,\dots,n} \\
			& \propto & \mathfrak{G}_{2,n} \H_2 \brackets{\braket{m}{+} \I_2 + \braket{m}{-} \Z_2} \ket{\psi}_2\ket{+}^{\otimes n-2}_{3,\dots,n},
\end{eqnarray*}
where the Hadamard operator $\H_i=(\X_i+\Z_i)/\sqrt{2}$. In other words, the quantum information has been teleported to the second qubit through the linear transformation 
\begin{equation}
\label{eq:telgate} \M = \H \brackets{\braket{m}{+} I + \braket{m}{-} \Z}.
\end{equation} 
Without loss of generality, the single-qubit state acting as the projector can be written as
\begin{equation}
\ket{m\brackets{\xi,\phi}} = \cos{\frac{\xi}{2}} \ket{+} + e^{i\phi} \sin{\frac{\xi}{2}} \ket{-}
\end{equation}
for some $0 \leq \xi < 2\pi$, $-\pi \leq \phi < \pi$. Thus, the linear transformation through which the quantum information is teleported can be written as
\begin{equation}
\M = \H \brackets{\cos{\frac{\xi}{2}} \I + e^{i\phi} \sin{\frac{\xi}{2}} \Z}.
\end{equation} 

In the special case that $\phi = \pm \pi$, corresponding to \ket{m} lying on the $x-y$ plane of the Bloch sphere, this transformation becomes the (familiar from cluster state MBQC) unitary transformation $\H\rz{\pm \xi}$. Thus, there is an entire single-parameter family of unitary gates through which the initial state \ket{\psi} can be teleported, each corresponding to a projection of the first qubit on to some state lying in the $x-y$ plane. This family is universal for single-qubit rotations: via four projections, corresponding to $\xi = 0,\xi_2,\xi_3,\xi_4$ respectively, one teleports the transformation
\begin{eqnarray}
\nonumber \U\brackets{\xi_2,\xi_3,\xi_4} & = & \H\rz{\xi_4}\H\rz{\xi_3}\H\rz{\xi_2}\H\rz{0} \\
\label{eq:udef}			& = & \rx{\xi_4}\rz{\xi_3}\rx{\xi_2}, 
\end{eqnarray}
which is an arbitrary single-qubit unitary decomposed in terms of Euler angles.

To this point, we have not discussed how to compensate for the randomness associated with the measurements. If one were to drive the gate teleportation described above via projective measurements, measurements must be made in an orthonormal basis containing \ket{m\brackets{\xi, \phi = \pi}}. For a single qubit, this basis would be $\mathcal{B} \brackets{\xi, \phi} = \set{\ket{m\brackets{\xi, \phi}}, \ket{m^{\perp}\brackets{\xi, \phi}}}$, where
\begin{eqnarray}
\label{eq:m} \ket{m\brackets{\xi, \phi}} & = &  \cos{\frac{\xi}{2}} \ket{+} + e^{i\phi} \sin{\frac{\xi}{2}} \ket{-}; \\
\label{eq:mperp} \ket{m^{\perp}\brackets{\xi, \phi}} & = & \sin{\frac{\xi}{2}} \ket{+} - e^{i\phi} \cos{\frac{\xi}{2}} \ket{-}.
\end{eqnarray}
From Eq.~(\ref{eq:telgate}), the teleported gate associated with the application of the projector \ketbra{m^{\perp}}{m^{\perp}} on the first qubit would be
\begin{eqnarray*}
\label{eq:telgateperp} \Mp & = & \H \sbrackets{\sin{\frac{\xi}{2}} \I - e^{i \phi} \cos{\frac{\xi}{2}} \Z}\\
& \equiv & \X \H \sbrackets{\cos{\frac{\xi}{2}} \I-e^{-i\phi} \sin{\frac{\xi}{2}} \Z},
\end{eqnarray*}
where in the last step we have made use of the identity $\X \H = \H \Z$, and have dropped an unimportant overall phase. Once again considering the special case that $\phi = \pm \pi$, this reduces to 
\begin{eqnarray*}
\Mp & = & \X\H \rz{\pm \xi} = \X \M.
\end{eqnarray*}
The teleported gate can be summarized succinctly as follows: measuring in the basis \set{\ket{m}, \ket{m^{\perp}}} defined by Eqs.~(\ref{eq:m},\ref{eq:mperp}) with $\phi = \pi$, and denoting the measurement outcome by $m=0$ for state \ket{m} and $m=1$ for \ket{m^{\perp}}, then the teleported gate is $\X^m \H \rz{\xi}$. The operator $\X$ can be thought of as a byproduct operator that occurs as a result of obtaining measurement outcome $1$. 

If the aim is to teleport the operator $\U\brackets{\xi_2,\xi_3,\xi_4}$ defined in Eq.~(\ref{eq:udef}), then apparently one runs into a problem should a measurement outcome of $1$ be obtained for any of the four measurements needed to teleport this gate. In fact the $\X$ byproduct operators can be pushed through the rotations because $\rz{\xi} \X = \X \rz{-\xi}$. Suppose then that one performs four projective measurements on a one-dimensional cluster state, with the $i$th measurement being a projective measurement of qubit $i$ in the orthonormal basis $\mathcal{B}\brackets{\theta_i,\pi}$, $\theta_1 = 0$ and the measurement outcome denoted $m_i \in \set{0,1}$. The quantum information originally situated on qubit 1 before the global entangling operation is then teleported through the gate
\begin{eqnarray}
\nonumber \M & = & \X^{m_4}\H\rz{\theta_4} \X^{m_3}\H\rz{\theta_3} \X^{m_2}\H\rz{\theta_2}\X^{m_1}\H \\
\nonumber 			& = & \X^{m_4} \Z^{m_3} \X^{m_2} \Z^{m_1} \rxs{\brackets{-1}^{m_3+m_1}\theta_4} \\
\label{eq:telmeasdef}	
&\times & \rzs{\brackets{-1}^{m_2}\theta_3} \rxs{\brackets{-1}^{m_1}\theta_2}.
\end{eqnarray}
Comparing Eq.~(\ref{eq:udef}) and Eq.~(\ref{eq:telmeasdef}), the choices $\theta_2 = \brackets{-1}^{m_1} \xi_2$, 
$\theta_3 = \brackets{-1}^{m_2} \xi_3$, and 
$\theta_4 = \brackets{-1}^{m_3+m_1} \xi_4$ make the implemented teleported gate
\begin{equation}
\M = \X^{m_4} \Z^{m_3} \X^{m_2} \Z^{m_1} \U\brackets{\xi_2,\xi_3,\xi_4}.
\end{equation}
Thus, any gate can be implemented by conditioning each of the last three measurement bases on the results of previous measurements, up to an overall Pauli byproduct operation. The byproduct is of no concern, as $\Z$ has no effect on computational basis states while $\X$ merely swaps them; this means that the effect of the byproduct can be taken into account simply by appropriate reinterpretation of the final measurement outcomes of the circuit, contingent on the intermediate measurement outcomes.

An equivalent description of this universal gate teleportation can be obtained
within the Matrix-Product State (MPS) representation~\cite{Gross2007,Gross2007a}
of the one-dimensional cluster state:
\begin{equation}
\ket{\textrm{Cl}_n}=\displaystyle \sum_{\vec{i}} A^{[1]}[i_1] A^{[2]}[i_2] \dots A^{[n]}[i_n] \ket{i_1 i_2 \dots i_n},
\label{MPSCl}
\end{equation}
where $\vec{i}$ is an n-bit string and the site matrices $\set{A^{[j]}[i_j]}$ 
are all two-by-two, except for the boundaries; the $\set{A^{[1]}[i_1]}$ are row 
vectors and the $\set{A^{[n]}[i_n]}$ are column vectors. The site matrices are 
not unique, but it is particularly convenient if they are chosen to satisfy
the relation $\sum_{i_j} A^{[j]}[i_j]A^{[j] \dag}[i_j] = I$ for each $j$,
corresponding to the canonical form of the MPS~\cite{Perez-Garcia2007}. For the
left and right boundaries one obtains $A^{[1]}[0] = \frac{1}{\sqrt{2}}\bra{+}$,
$A^{[1]}[1] = \frac{1}{\sqrt{2}}\bra{-}$, $A^{[n]}[0] = \ket{0}$, and 
$A^{[n]}[1] = \ket{1}$; for the bulk sites $1 < j < n$ they are
$A^{[j]}[0]=\frac{1}{\sqrt{2}}\H$ and $A^{[j]}[1]=\frac{1}{\sqrt{2}}\H\Z
=\frac{1}{\sqrt{2}}\X\H$. The `always-on' operator H is teleported on each 
measurement of a qubit, and the X gate serves as the byproduct operator. In
general, an MPS state is a universal resource for measurement-based 
single-qubit gate teleportation (a `computational wire') if the bulk site 
matrices can be chosen to be proportional to unitaries~\cite{Gross2010}. In 
this case they can be written as $A^{[j]}[0]=\frac{1}{\sqrt{2}}W$ and 
$A^{[j]}[1]=\frac{1}{\sqrt{2}}W\R_z(\phi)$ with $W\in SU(2)$ and
$\phi\in\mathbb{R}$.

In the context of the calculations presented in the next two sections, it is worth pointing out that there are two special features associated with the projective single-qubit measurements on one-dimensional cluster states presented above. The first is that the linear transformation $M$ on the quantum state \ket{\psi} is guaranteed to be unitary; in practice, this means that the effect of such a measurement is not dependent on the input state \ket{\psi}. If the linear transformation is not unitary (for example projections of the local system outside the $x-y$ plane, as discussed below), then 
%(assuming that the system comprising the cluster state and the measuring device is closed) 
there is an equivalent unitary transformation resulting in the same final state vector; however, the equivalent unitary will depend on \ket{\psi}. The second feature is that the byproduct operator resulting from measurement outcome $1$ is always $\X$. This is beneficial as $\X$ operators can be pushed through $\mathrm{R_z}$ operators with an easily characterized effect, as discussed above. 
These properties can be summarized as follows:
\begin{itemize}
\item \textbf{Property IA:} the teleported gate is in general of the form $\H \rz{\xi}$ where $\xi \in \mathbb{R}$, i.e. a unitary gate corresponding to a $z$-axis rotation by a real angle, followed by a Hadamard gate.
\item \textbf{Property IIA:} the byproduct operator is always $\X \equiv \rx{\pi}$.
\end{itemize}

The above two features do not hold for single-qubit projective measurements outside the $x-y$ plane. In general,
\begin{itemize}
\item \textbf{Property IB:} the teleported gate is in general of the form $\H \rz{\xi}$ where $\xi \in \mathbb{C}$, i.e. a non-unitary gate corresponding to a $z$-axis rotation by a complex angle, followed by a Hadamard gate.
\item \textbf{Property IIB:} the byproduct operator is in general $\rx{\eta}$ where $\eta \neq \pi$ in general, i.e. an $x$-axis rotation not corresponding simply to Pauli $\X$.
\end{itemize}
Another way to view this is that in $x-y$ plane, one always teleports $\H \rz{\xi + \delta_{m,1} \pi}$ with $\xi \in \mathbb{R}$, whereas in any other plane, one teleports $\H \rz{\xi + \delta_{m,1} \epsilon}$ where $\xi \in \mathbb{R}$ corresponds to the angle of the z-rotation in the desired gate, and $\epsilon \in \mathbb{C}$ is a complex error that occurs on measurement outcome 1.

Single-qubit gates alone are not sufficient for universal computation; at least one multiqubit entangling gate is required as well. In the cluster state model, multiqubit gates are accomplished via measurement patterns on 2D structures. Logical qubits are processed by horizontal 1D wires, while entangling operations are mediated by vertical links between them. An entangling gate that is locally equivalent to controlled-NOT can be achieved by measuring a link qubit in the $\Y$ basis~\cite{Raussendorf2001,Raussendorf2003}.

As with the case of single-qubit rotations, local Pauli byproduct operators may exist as well, depending on the measurement outcomes. As before, these byproduct operators are of no concern computationally. Thus there exists for cluster states a measurement pattern that deterministically implements a two-qubit unitary entangling gate, with any byproducts that occur being of the local Pauli type. It is not immediately clear that this will be the case for states other than the cluster state; in general, the teleported two-qubit gate may be non-unitary and the byproduct may be non-local. Any candidate resource state for MBQC must be shown to be amenable to a measurement pattern implementing some suitable two-qubit entangling gate. In Section~\ref{sec:results}, we describe some 1D structures that are resources for single-qubit rotations and then in the examples of Section~\ref{sec:randomlength}, we demonstrate how to perform entangling gates with natural 2D or 3D extensions of the 1D structures, and how to compensate the randomness associated the measurements.

\section{Projective measurements on SLOCC-transformed cluster states}
\label{sec:results}
As discussed in the previous section, the distinguishing feature of MBQC with regular cluster states is that there exists a plane of the Bloch sphere onto which successive, adaptive, single-qubit projective measurements drive an arbitrary computation that is deterministic and of fixed length. No matter which single-qubit gate is desired, it will be implemented with certainty up to an unimportant Pauli byproduct with four measurements. For an SLOCC-transformed cluster state, it is not obvious that there exists any such plane: in general it is not possible to simultaneously satisfy both Properties IA and IIA (or for the latter any another convenient Clifford gate). A natural question to ask is then: under what circumstances can either property IA or IIA be satisfied by itself? And if only one property is satisfied, does there remain a deterministic protocol for universal quantum computation? Sec.~\ref{subsec:strategyI} and Sec.~\ref{subsec:strategyII} discuss the circumstances under which it is possible to independently satisfy Property IA and IIA, respectively.

\subsection{Strategy I: Guaranteed Unitary Evolution}
\label{subsec:strategyI} 

\subsubsection{Derivation of N-type Operators}

For convenience, define
\begin{equation}
\mathbf{S}_{k,l} := \displaystyle \bigotimes_{j=k}^l S^{(j)}_j
\end{equation}
where $S^{(j)} \in \gl$. From Eq.~(\ref{eq:clusteridentity}), it is clear that the SLOCC-transformed cluster state encoding quantum information can be written in the form
\begin{equation}
\mathbf{S}_{1,n} \cl{n}' = \mathbf{S}_{2,n} \mathfrak{G}_{2,n} \brackets{aS^{(1)}_1\ket{0}_1 \I_2 + bS^{(1)}_1\ket{1}_1 \Z_2} \ket{+}^{\otimes n-1}_{2,\dots,n}.
\end{equation}
Following the procedure discussed in Sec.~\ref{sec:background}, applying the projector \ketbra{m}{m} to the first qubit yields the resulting state on the remaining qubits:
\begin{eqnarray*}
\ket{\Phi} & = & \mathbf{S}_{2,n} \mathfrak{G}_{2,n} \brackets{a\bra{m}S^{(1)}_1\ket{0} \I_2 + b\bra{m}S^{(1)}_1\ket{1} \Z_2} \ket{+}^{\otimes n-1}_{2,\dots,n} \\
		& = & \mathbf{S}_{2,n} \mathfrak{G}_{2,n}  \H_2 \M_2
		\ket{\psi}_2\ket{+}^{\otimes n-2}_{3,\dots,n},
\end{eqnarray*}
where
\begin{equation}
\label{eq:stelgate} \M_2 = \H_2\sbrackets{\frac{\bra{m}S^{(1)}\ket{+}}{\sqrt{2}} \I_2 + \frac{\bra{m}S^{(1)}\ket{-}}{\sqrt{2}}\Z_2}.
\end{equation}
The only way for this to correspond to a unitary gate is if
\begin{eqnarray*}
\frac{1}{\sqrt{2}}\bra{m}S^{(1)}\ket{+} & = & e^{i\alpha} \cos{\frac{\xi}{2}}; \\
\frac{1}{\sqrt{2}}\bra{m}S^{(1)}\ket{-} & = & -i e^{i\alpha} \sin{\frac{\xi}{2}},
\end{eqnarray*}
where $0 \leq \xi < 2\pi$, and 
therefore
\begin{equation}
\label{eq:mcond} S^{(1)\dagger}\ket{m} = \sqrt{2} e^{-i\alpha}\sbrackets{\cos{\frac{\xi}{2}}\ket{+} + i \sin{\frac{\xi}{2}}\ket{-}},
\end{equation}
or equivalently
\begin{eqnarray}
\ket{m} & = & \sqrt{2} \brackets{S^{(1)\dagger}}^{-1} e^{-i\alpha}\sbrackets{\cos{\frac{\xi}{2}}\ket{+} + i \sin{\frac{\xi}{2}}\ket{-}}\nonumber \\
& = & e^{-i\alpha} \brackets{S^{(1)\dagger}}^{-1} \rz{\xi} \ket{+}.
\label{eq:mcondwiths} 
\end{eqnarray}
Eq.~(\ref{eq:mcondwiths}) is the condition on the state \ket{m} such that measurement outcome 0 yields a unitary teleported gate. Note that there is a family of states characterized by a single parameter $\xi$ fulfilling this condition, not including the unimportant overall phase $\alpha$. 

To ensure that the measurement yields a unitary teleported gate independent of the measurement outcome, a similar condition must follow for the orthogonal complement \ket{m^{\perp}}. Orthogonality requires
\begin{equation}
\bra{m^{\perp}} \propto \bra{-} \rz{-\xi} S^{(1)\dagger},
\end{equation}
and therefore
\begin{equation}
\label{eq:mperpwiths} \ket{m^{\perp}} = c S^{(1)} \rz{\xi} \ket{-}
\end{equation}
for some constant $c \in \mathbb{C}$. Repeating the procedure that led to Eq.~(\ref{eq:mcond}), but with \ket{m^{\perp}} instead of \ket{m}, one obtains
\begin{equation}
\label{eq:mperpcond} S^{(1)\dagger}\ket{m^{\perp}} = \sqrt{2} e^{-i\beta}\sbrackets{\cos{\frac{\xi}{2}}\ket{+} + i \sin{\frac{\xi}{2}}\ket{-}}.
\end{equation}
Substituting Eq.~(\ref{eq:mperpwiths}) into Eq.~(\ref{eq:mperpcond}) yields
\begin{eqnarray}
\nonumber &&\sqrt{2} e^{-i\beta}\sbrackets{\cos{\frac{\xi}{2}}\ket{+} + i \sin{\frac{\xi}{2}}\ket{-}} = c S^{(1)\dagger} S^{(1)} \rz{\xi} \ket{-} \\
\label{eq:last} & = & c S^{(1)\dagger} S^{(1)} \sbrackets{\cos{\brackets{\frac{\xi}{2}}} \ket{-} - i \sin \brackets{\frac{\xi}{2}} \ket{+}}.
\end{eqnarray}
Rewriting Eq.~(\ref{eq:last}) in the computational basis results in the expression
\begin{equation}
\label{eq:keyforunitarycond} \ket{0} + e^{-i \xi}\ket{1} = c^{\prime} S^{(1)\dagger} S^{(1)} \brackets{\ket{0} - e^{i \xi} \ket{1}}
\end{equation}
for a suitably defined constant $c^{\prime}$. Defining $T_k := S^{(k)\dagger}S^{(k)}$ and $T_{k}^{i,j} := \bra{i} T_k \ket{j}$, we can see from Eq.~(\ref{eq:keyforunitarycond}) that
\begin{eqnarray}
c^{\prime} \brackets{T_{1}^{0,0} - e^{i \xi} T_{1}^{0,1}} & = & 1;\\
c^{\prime} \brackets{T_{1}^{1,0} - e^{i \xi} T_{1}^{1,1}} & = & e^{-i \xi}.
\end{eqnarray}
From here, it is easily deduced that
\begin{equation}
\left | T_{1}^{0,0} - e^{i \xi} T_{1}^{0,1} \right | = \left | T_{1}^{1,1} 
- e^{-i \xi} T_{1}^{1,0} \right |.
\end{equation}
Making use of the Hermiticity of $T_k$, simple algebra yields
\begin{equation}
\label{eq:unitarycond} T_{1}^{0,0} = T_{1}^{1,1}.
\end{equation}
Eq.~(\ref{eq:unitarycond}) above has a very simple geometric interpretation: it means that $S^{(1)}$ must preserve the relative norm of the computational basis states. There is no requirement for $S^{(1)}$ to preserve their orthogonality, however, which means that $S^{(1)}$ is allowed to differ quite drastically from a unitary transformation; in fact, it can be made arbitrarily close to singular, as the relative angle between the computational basis states under the transformation by $S^{(1)}$ can be vanishingly small. We will refer to operators obeying this norm-preservation restriction as $\mathrm{N}$-type operators.

\begin{defn}
A $\gl$ operator $S$ satisfying $\bra{0}S^{\dagger}S\ket{0} = \bra{1}S^{\dagger}S\ket{1}$ is called an \N-type operator.
\end{defn}

%Without loss of generality, we can write the local $\gl$ transformations $\mathrm{S}^{(i)}$ as
%\begin{eqnarray*}
%	\mathrm{S}^{(i)} & = & \sqrt{2} \mathrm{diag}(\cos{(\theta^{(i)})}, \sin{(\theta^{(i)})}) \\
%				 & & \times \rz{\alpha^{(i)}} \rx{\beta^{(i)}} \rz{\gamma^{(i)}}, 
%\end{eqnarray*}
%where  $0 < \theta^{(i)} < \frac{\pi}{2}$ and $-\pi \leq \alpha^{(i)}, \beta^{(i)}, \gamma^{(i)} < \pi$.

The singular value decomposition is helpful for characterizing \N-type operators. An arbitrary SLOCC operator $S$ can be written in terms of its singular value decomposition as $S = \U D\V$, where $\U$ is an arbitrary two-qubit unitary, 
$D$ is a positive-definite diagonal matrix 
\begin{equation}
\label{eq:D} D = \kappa \mtxtwotwo{\cos{\theta}}{0}{0}{\sin{\theta}},
\end{equation}
where $0 < \theta < \frac{\pi}{2}$, and $\V$ is an arbitrary unitary matrix 
parametrized via the Euler decomposition as 
$\V = \H \rz{\alpha} \rx{\beta} \rz{\gamma}$, 
$0 \leq \alpha, \beta, \gamma < 2\pi$, and any global phase has been absorbed 
into $\U$. It is then straightforward to determine that
\begin{eqnarray}
\bra{0}S^{\dagger}S\ket{0} &=& \frac{1}{2} \kappa^2 \brackets{1+\cos{2\theta}\sin{\alpha}\sin{\beta}};\\
\bra{1}S^{\dagger}S\ket{1} &=& \frac{1}{2} \kappa^2 \brackets{1-\cos{2\theta}\sin{\alpha}\sin{\beta}}. 
\end{eqnarray}
So, if $S$ is an \N-type operator, we must have $\theta=\frac{\pi}{4}$ (in which case $S$ is proportional to a unitary), $\alpha=0$ or $\beta=0$. The case where $\beta=0$ still allows us to assume $\alpha=0$ without loss of generality. Doing so, the $\rx{\beta}$ operator can be commuted past the $\H$ to turn into a $z$-rotation, and then absorbed into $\U$. Thus we have the following characterization of \N-type operators.

\begin{lem}
\label{lem:ntype} Every \N-type operator $S$ must either be proportional to a unitary operator, or of the form $S=\U D\V$, where $\U$ is an arbitrary two-by-two unitary operator, $D$ is defined as in Eq.~(\ref{eq:D}) and $\V = \H\rz{\gamma}$ with $0 \leq \gamma < 2\pi$. 
\end{lem}

It is straightforward to obtain an expression for the byproduct angle $\mu'$ in the case of measurement outcome 1 in terms of the parameters $\theta$, $\gamma$ and $\xi$ (recall that this is the degree of freedom in the measurement basis). Using Eq.~(\ref{eq:stelgate}), it can easily be checked that the teleported gate when $\ket{m} \propto \brackets{S^{\dag}}^{-1}\rz{-\xi}\H\ket{0}$ is $\M=\H\rz{\xi}$, and when $\ket{m} \propto S \rz{-\xi}\H\ket{1}$ is $\Mp=\rx{\mu^{\prime}}\H\rz{\xi}$, where the byproduct angle $\mu^{\prime}$ obeys
\begin{equation}
\label{eq:byproductangle} \tan{\frac{\mu^{\prime}}{2}} =  \frac{1-\cos{2\theta}\cos{(\gamma-\xi)}}{\cos{2\theta}\sin{(\gamma-\xi)}}.
\end{equation}
The probabilities of the two measurement outcomes can also be easily calculated in terms of the same parameters, and are found to be
\begin{eqnarray}
\label{eq:zeroprobs1} p(0) & = & \frac{1}{2}\brackets{1+\cos{2\theta}\cos{2\xi}};\\
\label{eq:oneprobs1} p(1) & = & \frac{1}{2}\brackets{1-\cos{2\theta}\cos{2\xi}}.
\end{eqnarray}

This differs from the case of gate teleportation with a perfect cluster state, where the two measurement probabilities are always exactly $\frac{1}{2}$. That said, the expected probability of obtaining a byproduct here, averaged over all $\xi$, is
\begin{equation}
\langle p(1) \rangle_{\xi} = \frac{1}{2},
\end{equation}
irrespective of $\theta$. We can thus generically expect to obtain an unwanted byproduct operator that we must compensate on half of our single-qubit measurements. This point will be discussed in Sec.~\ref{sec:randomlength}.

\subsubsection{Properties of N-Transformed Cluster States}
\label{sec:ntransformed}

Cluster states locally transformed by $\N$-type operators can exhibit remarkably different properties from perfect cluster states. Nevertheless, as will be shown later in Sec.~\ref{sec:2DNUN}, they can under some circumstances serve as universal resources for MBQC of random length. 

Consider for the moment the Schmidt decomposition of an $n$-qubit cluster state on some set of qubits $\mathcal{V}$ with respect to a bipartition separating qubit $k$ from the rest:
\begin{equation}
\label{eq:schmidt} \cl{n}_{\mathcal{V}} = \frac{1}{\sqrt{2}} \brackets{\ket{0}_k \ket{\textrm{Cl}_{n-1}}_{\mathcal{V} \setminus k}+\ket{1}_k \Z_{\mathcal{N}(k)} \ket{\textrm{Cl}_{n-1}}_{\mathcal{V} \setminus k}},
\end{equation}
where $\ket{\textrm{Cl}_{n-1}}_{\mathcal{V} \setminus k}$ refers to the cluster state resulting from deleting qubit $k$ and $\Z_{\mathcal{N}(k)}$ is the tensor product of $\Z$ operators acting on all the neighbors of $k$. This can be checked by verifying that this state satisfies the stabilizer conditions~(\ref{eq:stabconds}). The Schmidt basis for the multiqubit component can be further decomposed if desired by the same technique. The equality of the Schmidt coefficients in Eq.~(\ref{eq:schmidt}) demonstrates that any individual qubit in a cluster state has a maximally mixed local reduced density matrix, or in other words that it is maximally entangled with the rest of the cluster with respect to the von Neumann entanglement entropy. Likewise, exactly one ebit of entanglement is shared across any bipartition of the cluster state.

One effect of $\N$-type operators is to change the local reduced density 
matrices of individual qubits within the state. In the canonical 
representation, the site matrices of the MPS representation [cf.\ 
Eq.~(\ref{MPSCl})] are 
$A^{[j]}[0]=\frac{1}{\sqrt{2}}W=\H\rz{-(\gamma^{(j)}+2\theta^{(j)})}$ and 
$A^{[j]}[1]=\frac{1}{\sqrt{2}}W\rz{4\theta^{(j)}}$, which are both unitary.
This immediately implies that the channel having these matrices as Kraus operators is unital, and like the ordinary 
cluster state one ebit of entanglement is shared across any bipartition. That 
said, the entanglement between any given qubit and the rest of the system need 
not be unity.

Consider for example the local reduced density matrix of a qubit adjacent to an endpoint of a 1D cluster state with $n$ qubits, numbered 1 to $n$ from left to right. Singling out first qubit 2 and then qubit 3, this state can be written as
\begin{equation}
\ket{\textrm{Cl}^{1D}_n}_{1\dots n}=\frac{1}{\sqrt{2}}\left( \ket{0}_2 \ket{+}_1
I_3 +\ket{1}_2 \ket{-}_1 \Z_3\right) \ket{\textrm{Cl}^{1D}_{n-2}}_{3 \dots n}.
\label{eq:schmidt1D} 
\end{equation}
Now consider the action of an $\N$-type operator, $N^{(2)} = D\H\rz{\gamma}$ on qubit 2, where the leading $U$ operator is dropped because it can be absorbed into the measurement basis. It is easy to check that 
\begin{equation}
N^{(2)}\ket{\textrm{Cl}^{1D}_n}_{1\dots n} = \frac{1}{\sqrt{2}}\left(\cos{\theta} \ket{0}_2 \ket{\Phi} - i\sin{\theta}\ket{1}_2 \ket{\Phi^{\perp}}\right),
\label{eq:nschmidt1D}
\end{equation}
where $\ket{\Phi}$ and $\ket{\Phi^{\perp}}$ are $\gamma$-dependent states for qubits $1,3,\ldots,n$ such that $\braket{\Phi}{\Phi^{\perp}}=0$. Thus, Eq.~(\ref{eq:nschmidt1D}) remains a Schmidt decomposition. The local reduced density matrix of qubit 2 is
\begin{equation}
\nonumber \rho_2 = \mtxtwotwo{\mathrm{cos}^2\theta}{0}{0}{\mathrm{sin}^2\theta},
\end{equation}
revealing that qubit 2 is no longer maximally entangled with the rest of the state. A similar calculation can be performed for qubits further from the boundary, with qualitatively similar results.

%This argument does not easily generalize to qubits in different positions in the chain. Because $\N$-type operators are not diagonalizable they do not preserve the orthogonality of any single-qubit orthogonal basis. Consequently, if the qubit of interest is not adjacent to an endpoint of the chain, the Schmidt decomposition of the transformed state is non-trivial to obtain. Nevertheless, this simple example demonstrates that $\N$-type operators have a non-trivial effect on the local reduced density matrices. 

Another property of these states is the long-range behavior of two-point correlation functions, those of the form $C_{i,j}(\mathcal{A},\mathcal{B}) := \langle \mathcal{A}_i \mathcal{B}_j \rangle - \langle \mathcal{A}_i \rangle \langle \mathcal{B}_j \rangle$ for some operators $\mathcal{A}$ and $\mathcal{B}$. Two-point correlation functions of large 1D cluster states with periodic boundary conditions can be efficiently calculated using the Matrix Product State (MPS) representation. For an $n$-qubit ring, calculation of the correlation functions amounts to taking traces of products of $n$ 4 $\times$ 4-dimensional matrices. Consider therefore a 1D cluster state with periodic boundary conditions (i.e.\ a ring), with the operation $N^{(i)}$ acting on qubit $i$. For this state, calculations show that all two-point Pauli correlation functions vanish except for the second-nearest-neighbour correlation function $C_{i-1,i+1}(Z,Z) = \cos{2\theta} \sin{\gamma}$. This is in contrast to the perfect cluster state with periodic boundary conditions, for which all two-point correlation functions identically vanish.

\begin{figure}
\includegraphics[scale=0.9]{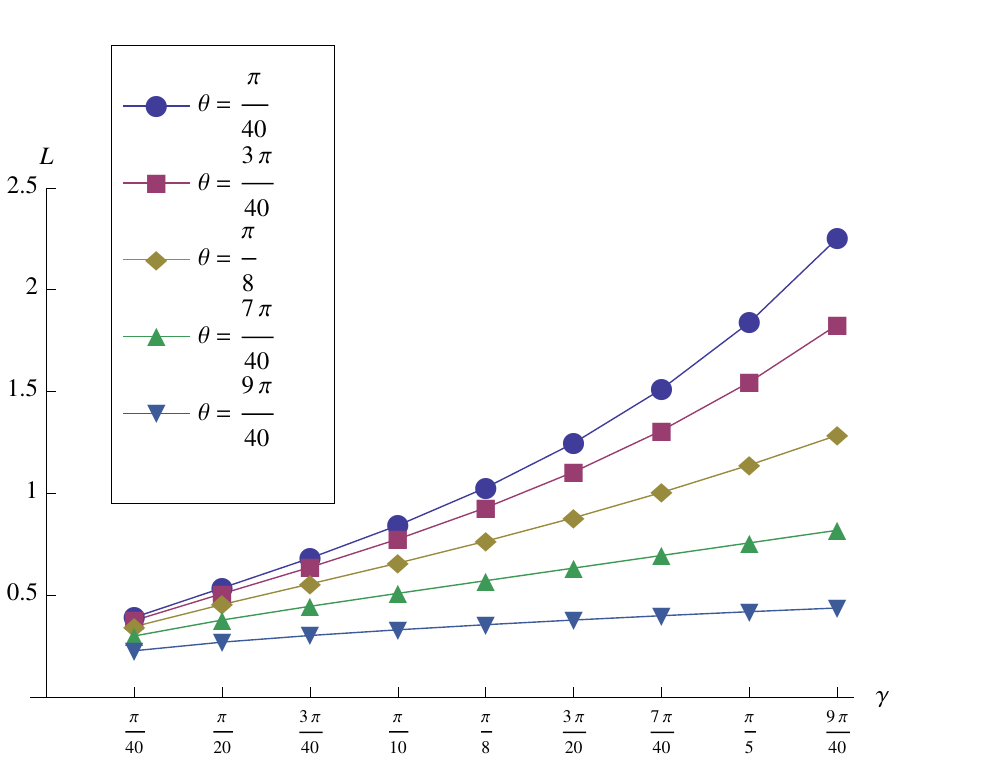}
\caption{(Color online) Correlation length scale $L$ associated with the correlation function $\abs{C_{1,2j+1}(Z,Z)} \sim\exp\brackets{-\frac{2j}{L}}$ for a $\N-\U-\N$ ring with all $\N$-type operators identical, on a ring of 1000 qubits, as a function of the parameters of $\gamma$ and $\theta$, the parameters of $N$. The length scale increases as $N$ approaches the singular limit, i.e.\ as $\theta$ gets close to $0$ or $\pi/2$.}
\label{fig:exponents}
\end{figure}

As another example, the relevance of which will become clear in Sec.~\ref{ex:exam1}, consider a ring with an even number of qubits, with $N$ acting on every alternate qubit; say, the ones with even labels. In this case, the magnitude of the same two-point correlation function between odd-numbered qubits decays exponentially:
\begin{equation}
\abs{C_{1,2j+1}(Z,Z)} \sim\exp\brackets{-\frac{2j}{L}}.
\end{equation}
The length scale $L$ depends on $\theta$ and $\gamma$. The same correlation function between pairs of qubits with at least one even label is zero. The numerically obtained behavior of $L$ for a ring of 1000 qubits is shown in Fig.~\ref{fig:exponents} as a function of $\gamma$ for several values of $\theta$ between $0$ and $\pi/4$. As can be seen from the figure, the length scale increases with decreasing $\theta$ over this range, i.e. as the $\N$-type operators approach the singular limit $\theta=0$. The length scale is symmetric about $\theta = \pi/4$ between $0$ and $\pi/2$. Viewed as a function of $\gamma$ with $\theta$ held constant, the correlation function is convex and non-negative in $\gamma$ over the interval from $0$ to $\pi/2$ and is symmetric about $\pi/4$. For $\pi/2 \leq \gamma < \pi$, the magnitudes behave the same way as in the previous interval, but the signs alternate. The $\gamma$-behavior is periodic with period $\pi$. We note in passing that these non-zero correlation functions provides a lower bound for the localizable entanglement~\cite{Popp2005} between that pair of qubits in the state via projective measurements, with respect to the concurrence~\cite{Wootters2001}. 

Note that a number of resources for MBQC with non-vanishing long-range correlation functions have been pointed out in the literature~\cite{Gross2007a,Brennen2008,Miyake2010,Bartlett2010,Wei2011}, based on the so-called spin-1 AKLT model~\cite{Affleck1987,Affleck1988}. These states are quasi-deterministic resources, in the sense that measurement-based computations using these states can be made arbitrarily likely to succeed, either by reduction of the resource state to a deterministic resource or by a repeat-until-success strategy with each elementary gate requiring a random number of measurements. In Secs.~\ref{ex:exam1} and~\ref{sec:2DNUN}, we describe resource states called $\N-\U-\N$ states that are based on cluster states transformed by $\N$-type operators; these states share the properties of quasi-determinism and non-vanishing long-range correlations.

%Finally, from exact calculations on non-periodic chains of up to 7 qubits with $\N$ operators acting on even qubits, the entropy of entanglement with respect to any bipartition separating the left of the chain from the right appears to be 1. This feature is also true of the 1D cluster state. The entropy of entanglement with respect to any other bipartition is not maximal, however, which implies that the state is not LU-equivalent to a cluster. Nevertheless, the fact that the left and right halves of such a chain share a full ebit of entanglement suggests that they may be useful for processing a single logical qubit. In Sec.~\ref{ex:exam1}, we show that this is indeed the case.

%For MBQC, we must be able to undo the effect of this non-Pauli byproduct operator, which is possible if every $\mathrm{S}^{(i)}$ operator for $i$ even is unitary.
\subsection{Strategy II: Guaranteed Pauli Byproduct}
\label{subsec:strategyII}

\subsubsection{Derivation of B-Type Operators}

Another possible strategy is to attempt to ensure that the byproduct operator is guaranteed to be Pauli-$\X$, whether or not the teleported linear transformation is unitary. The advantage of this approach is that $\X$ has nice commutation properties through rotation operators about the $z$-axis, whether they be by real or complex angles, leading to the hope that the randomness inherent in the measurement process can be easily compensated. 

When projectively measuring in the orthonormal basis \set{\ket{m}, 
\ket{m^{\perp}}}, the two possible operations that can occur are
\begin{eqnarray}
 \label{eq:nobyproduct} \M & = & \H \sbrackets{\frac{\bra{m}S^{(1)}\ket{+}}{\sqrt{2}}\I + \frac{\bra{m}S^{(1)}\ket{-}}{\sqrt{2}}\Z};\\
 \Mp & = & \H \sbrackets{\frac{\bra{m^{\perp}}S^{(1)}\ket{+}}{\sqrt{2}}\I + \frac{\bra{m^{\perp}}S^{(1)}\ket{-}}{\sqrt{2}}\Z}\\
		& = & \X \H \sbrackets{\frac{\bra{m^{\perp}}S^{(1)}\ket{-}}{\sqrt{2}}\I+\frac{\bra{m^{\perp}}S^{(1)}\ket{+}}{\sqrt{2}}\Z}.
\end{eqnarray}
Since $\I$ and $\Z$ are linearly independent, it follows that for the byproduct to be guaranteed to be proportional to Pauli-$\X$, one must have
\begin{eqnarray}
\label{eq:xbycond11} \bra{m}S\ket{+} & = & c \bra{m^{\perp}}S\ket{-}; \\
\label{eq:xbycond21} \bra{m}S\ket{-} & = & c \bra{m^{\perp}}S\ket{+},
\end{eqnarray}
or equivalently,
\begin{eqnarray}
\label{eq:xbycond12} \bra{m}S\ket{0} & = & c \bra{m^{\perp}}S\ket{0}; \\
\label{eq:xbycond22} \bra{m}S\ket{1} & = & -c \bra{m^{\perp}}S\ket{1},
\end{eqnarray}
for some non-zero constant $c \in \field{C}$. Suppose $S=\U D\V$ where $\U$ is 
an arbitrary single-qubit unitary, $D$ is defined in Eq.~(\ref{eq:D}), and 
$V=\rz{\beta}\rx{\gamma}\rz{\delta}$. Further suppose that $\ket{m}=UU^{\prime}\ket{0}$ and $\ket{m^{\perp}}=UU^{\prime}\ket{1}$, with $U^{\prime}=\rz{\beta^{\p}}\rx{\gamma^{\p}}\rz{\delta^{\p}}$. The reason for the appearance of $U$ in the definitions of \ket{m} and \ket{m^{\perp}} is to compensate for the appearance of $U$ in the singular value decomposition of $S$. The only effect of the $\rz{\delta^{\p}}$ operation is to multiply the teleported gate by a global phase, so we can choose $\delta^{\p}=0$ in what follows without loss of generality (it remains a free parameter for the applied unitary $U'$). Having done so, 
Eqs.~(\ref{eq:xbycond12}-\ref{eq:xbycond22}) can be rewritten as
\begin{eqnarray}
\label{eq:xbycond13} \bra{0}Q\ket{0} & = & c \bra{1}Q\ket{0}; \\
\label{eq:xbycond23} \bra{0}Q\ket{1} & = & -c \bra{1}Q\ket{1},
\end{eqnarray}
where we have defined
\begin{eqnarray}
\label{eq:qdef} Q &:=& (U^{\p})^{\dagger}DV \\
\nonumber			    & = & \rx{-\gamma^{\p}}\rz{-\beta^{\p}}D\rz{\beta}\rx{\gamma}\rz{\delta} \\
\nonumber			    & = & \rx{-\gamma^{\p}}\rz{\beta-\beta^{\p}}D\rx{\gamma}\rz{\delta} \\
\label{eq:qdeffinal}			    & := & \rx{-\gamma^{\p}}\rz{b}D\rx{\gamma}\rz{\delta};
\end{eqnarray}
in the last line above we have defined $b := \beta - \beta^{\p}$. In the expression above, $\gamma^{\p}$ and $b$ are free parameters, while $D$, $\gamma$ and $\delta$ are determined by the SLOCC operator $S$. 

Return now to the constraints, Eqs,~(\ref{eq:xbycond13}-\ref{eq:xbycond23}). Denoting $Q_{ij} := \bra{i}Q\ket{j}$, one finds that $Q_{00}/Q_{10} = -Q_{01}/Q_{11}$. Note that neither $Q_{10}$ nor $Q_{11}$ can be zero; if either were zero, then the constraints would force $Q$ and therefore $S$ to be singular, which by assumption is not the case. This in turn means that 
\begin{equation}
\label{eq:qdetidentity} \textrm{Det}(Q) = 2Q_{00}Q_{11}. 
\end{equation}
From the definition of $Q$, Eq.~(\ref{eq:qdeffinal}), 
\begin{eqnarray}
\textrm{Det}(Q) & = & \sin{2\theta};\nonumber \\
\nonumber 2 Q_{00}Q_{11} & = & \sin{\gamma^{\p}}\sin{\gamma}\brackets{\cos{\gamma}-i\cos{2\theta}\sin{b}} \nonumber \\
			 &+& \sin{2\theta}\brackets{1+\cos{\gamma}\cos{\gamma^{\p}}}.
\end{eqnarray}
Substituting the above expressions into Eq.~(\ref{eq:qdetidentity}) and equating real and imaginary parts gives us the two conditions
\begin{eqnarray}
\label{eq:xbyrecond} \sin{\gamma^{\p}} \sin{\gamma} \sin{b} \cos{2\theta} &=& 0; \\
\label{eq:xbyimcond} \cos{\gamma}\brackets{\sin{\gamma^{\p}}\sin{\gamma}+\cos{\gamma^{\p}}\sin{2\theta}} & = & 0.
\end{eqnarray}
Recall that since $S$ is invertible, we cannot have $\sin{2\theta}=0$ and since $S$ is non-unitary, we cannot have $\cos{2\theta}=0$. The only ways to satisfy Eq.~(\ref{eq:xbyrecond}) are if $\sin{\gamma^{\p}}\sin{\gamma}=0$ or $\sin{b}=0$. In the first case, Eq.~(\ref{eq:xbyimcond}) immediately implies that $\cos{\gamma^{\p}}\cos{\gamma} = 0$, leaving $b$ as a free parameter for our measurement basis. In the second case, $\gamma^{\p}$ is fixed in terms of $\theta$ and $\gamma$, leaving no freedom in the measurement basis we are using. Furthermore, if we choose $\sin{\gamma^{\p}}=0$, i.e. $\gamma^{\p} \in \set{0,\pi}$, then the measurement basis we are using is restricted to being the computational basis acted on by $U$ (completely specified by $S$); again, no freedom. Therefore, the only solutions available to us that leave freedom in the measurement basis, and thus the teleported gate, are $\gamma \in \set{0,\pi}$ and $\gamma^{\p} \in \set{\frac{\pi}{2},\frac{3\pi}{2}}$. Note that
\begin{equation}
S^{\dagger}S = \mtxtwotwo{1+\cos{\gamma}\cos{2\theta}}{-ie^{i\delta}\sin{\gamma}\cos{2\theta}}{ie^{-i\delta}\sin{\gamma}\cos{2\theta}}{1-\cos{\gamma}\cos{2\theta}}.
\end{equation}
Thus, demanding that the SLOCC operators allow for a guaranteed Pauli by-product, assuming the SLOCC operator is not unitary and thus $\cos{2\theta} \neq 0$, is equivalent to demanding that $S^{\dagger}S$ be diagonal. Geometrically, this means that $S$ must preserve the overlap of the computational basis states (the transformed computational basis is still orthogonal). We will refer to this kind of basis-preserving operator as a $\mathrm{B}$-type operator. 

\begin{defn}
A $\gl$ operator $S$ satisfying $\bra{0}S^{\dagger}S\ket{1} = \bra{1}S^{\dagger}S\ket{0} = 0$ is called a $\mathrm{B}$-type operator.
\end{defn}

A B-type operator can therefore be written 
\begin{equation}
\nonumber \B=\begin{cases}
\U D\R_z[\beta]\R_z[\delta], & \gamma=0\\
\U D\R_z[\beta]\X\R_z[\delta], & \gamma=\pi,
\end{cases}
\end{equation}
ignoring overall phases. The two possibilities above can be simplified and collapsed into one. First, note that $\rz{\beta}$ can be commuted past $D$ and absorbed into $\U$. Next, note that $\X D \X$ is itself a diagonal matrix that results from swapping the diagonal entries of $D$. This means that the case where $\gamma =\pi$ can be written instead as $\U^{\prime} D^{\prime} \rz{\delta}$, where $\U^{\prime} = \U  \rz{\beta} \X$ and $D^{\prime} = \X D \X$. Of course, $\rz{\delta}$ can also be absorbed into $\U^{\prime}$; thus, a simple and completely general expression for a $\B$-type operator is
\begin{equation}
\label{eq:bchar} \B = \U D.
\end{equation}
The diagonal matrix $D$ in the singular value 
decomposition can be expressed as
$$D\propto{\rm diag}\left(\cos(\theta),\sin(\theta)\right)
=\sqrt{\sin(\theta)\cos(\theta)}\R_z[i\ln\cot(\theta)],$$
so that the \B-type operator becomes
\begin{equation}
\B \propto \U\R_z[i\ln\cot(\theta)].
\label{eq:BtypeRz}
\end{equation}
Because the unitary $\U$ can be absorbed directly into the measurement basis, one can interpret B-type 
operators as $z$-rotations by an imaginary angle, the value of which is related 
to the ratio of the singular values.

When the local operator is $\mathrm{B}$-type, the single-parameter family of measurement bases satisfies $\gamma^{\p} \in \set{\frac{\pi}{2},\frac{3\pi}{2}}$, and $\beta^{\p} \in [0,2\pi)$ is a free parameter. When this family of bases is used, the byproduct operator associated with measurement outcome 1 is always $\Z$ (up to a global phase). The teleported linear transformation is no longer unitary, however; it takes the form of a rotation about the $z$-axis of the Bloch sphere by a complex angle, followed by a Hadamard operation. The real part of the angle is completely specified by the choice of measurement basis, via the free parameter $\beta'$. The imaginary part is purely a function of the ratio of the singular values of the local $\gl$ operator. Denoting the measurement outcome corresponding to $\gamma^{\p} = \frac{\pi}{2}$ by $m=0$ and that for $\gamma^{\p} = \frac{3\pi}{2}$ by $m=1$, the teleported gate is given (up to a global phase) by

\begin{equation}
\label{eq:strategy2table}M = \X^m \H\rz{\beta^{\p}+i\textrm{ln}\cot{\theta}}.
\end{equation}

%Table~\ref{tab:strategy2table} summarizes the teleported transformation (up to an overall multiplicative factor) resulting from the different possible of combinations of $\gamma$ and $\gamma^{\p}$.

%can compensate for the angles $\beta$ and $\delta$ defining $S$
%\begin{table}
%\begin{tabular}{cccc}
%\hline
%$\gamma$ & $\gamma^{\p}$ & Teleported Transformation \\
%\hline
%$0$ & $\frac{\pi}{2}$ & \H\rz{\beta + \beta^{\prime} + \delta + \frac{\pi}{2} - i \mathrm{ln}\tan{\theta}} \\
%$0$ & $\frac{3\pi}{2}$ & \H\rz{\beta + \beta^{\prime} + \delta - \frac{\pi}{2} - i \mathrm{ln}\tan{\theta}} \\
%$\pi$ & $\frac{\pi}{2}$ & \H\rz{-(\beta + \beta^{\prime}) + \delta - \frac{\pi}{2} + i \mathrm{ln}\tan{\theta}} \\
%$\pi$ & $\frac{3\pi}{2}$ & \H\rz{-(\beta + \beta^{\prime}) + \delta + \frac{\pi}{2} + i \mathrm{ln}\tan{\theta}} \\
%\hline
%\end{tabular}
%\caption{Teleported linear transformation resulting from a measurement in the basis defined by a particular $\gamma^{\p}$, with $\beta^{\prime}$ a free real parameter, when the local $\gl$ operator is $\mathrm{B}$-type and defined by real parameters $\theta$, $\beta$, $\gamma$ and $\delta$. Here, $\tan{\theta}$ is the ratio of the singular values of the local $\B$-type operator.}
%\label{tab:strategy2table}
%\end{table}

\subsubsection{Properties of B-Transformed Cluster States}

Interpreting the B-type operators as $z$-rotations by imaginary angles 
provides a simple insight into the nature of B-transformed cluster states. The 
$\R_z$ operator commutes with 
all $\mathrm{CZ}$ gates, so one can push it all the way through to the 
$|+\rangle$ states in the definition of the cluster state, 
Eq.~(\ref{eq:defcluster}). Because $\R_z(\xi)|+\rangle$ is an arbitrary
single-qubit state, B-transformed cluster states are equivalent to applying 
$\mathrm{CZ}$ gates between qubits in arbitrary states (not including
computational basis states, which would require singular B operators).

One might assume that B-transformed cluster states are equivalent to weighted 
cluster states~\cite{Dur2005,Calsamiglia2005,Anders2006,Hein2006,Anders2007}, 
but this is not in fact the case. Weighted graph states are defined as 
$\prod_{\langle i,j\rangle}\mbox{CP}(\varphi)_{i,j} \ket{+}^{\otimes n}$,
where the controlled-phase entangling gate is $\mbox{CP}(\varphi)=\rm{diag}
\left(1,1,1,e^{i\varphi}\right)$; the cluster-state edge weights are then
given by $w_{ij}=\varphi_{ij}$. Consider the simplest counter-example of a
three-qubit linear cluster state with the central qubit transformed by a B-type
operator $\B=D\R_z[\gamma]$ with $D=\mbox{diag}(\cos\theta,\sin\theta)$. The 
eigenvalues of the local reduced density matrices are all 
$\set{\frac{1}{2}\brackets{1 \pm \cos{2\theta}}}$. On the other hand, for a 
three-qubit 1D weighted graph state with edge weights $\varphi_{12}$ and 
$\varphi_{23}$, the eigenvalues of the reduced density matrix are 
$\frac{1}{2}\brackets{1\pm \cos{\frac{\varphi_{12}}{2}}}$,
$\frac{1}{2}\brackets{1\pm \frac{1}{2}\cos{\varphi_{12}}\cos{\varphi_{23}}}$, 
and $\frac{1}{2}\brackets{1\pm \cos{\frac{\varphi_{23}}{2}}}$ for qubits 1
through 3, respectively. If the weighted graph and the B-transformed 
cluster are LU-equivalent, there must be some choice of $\varphi_{12}$ and 
$\varphi_{23}$ such that the spectra of the reduced density matrices are the 
same in both cases. For qubits 1 and 3 this implies $\phi_{12} = \phi_{23} 
= 4\theta$. For qubit 2 one obtains $\frac{1}{2}\brackets{1 \pm \frac{1}{2} 
\cos{4\theta}^2}$. This matches the corresponding spectrum for the 
B-transformed cluster only when $\theta = \pm \frac{\pi}{4}, \varphi = \pm \pi$,
in which case both states are LU-equivalent to a perfect cluster.

Cluster states locally transformed by $\B$-type operators also exhibit different properties from perfect cluster states. As with $\N$-type operators, $\B$-type operators change the local reduced density matrices of individual qubits within the state, as described in the following lemma. 

\begin{lem}
\label{lem:blocal}
Let $\ket{\mathrm{Cl}_n}_{\mathcal{V}}$ be an $n$-qubit cluster state on the set of qubits $\mathcal{V}$, with some subset $\mathcal{Q} \subseteq \mathcal{V}$ acted upon by $\B$-type operators. In particular, suppose that for each qubit $i \in \mathcal{Q}$, the \B-type operator acting is given by $B^{(i)} = D^{(i)}$ with $D^{(i)} = \sqrt{2}\mathrm{diag}\brackets{\cos{\theta^{(i)}},\sin{\theta^{(i)}}}$. Then, the local reduced density matrix for any qubit $k \in \mathcal{V}$ is given by
\begin{eqnarray*}
\rho_k & = & \mathrm{cos}^2\theta^{(k)}\ketbra{0}{0} + \mathrm{sin}^2\theta^{(k)}\ketbra{1}{1} \\ 
&+& \brackets{\frac{1}{2} \sin{2\theta^{(k)}}\displaystyle\prod_{j \in \mathcal{N}(k)}\cos{2\theta^{(j)}}\ketbra{0}{1} + \mathrm{h.c.}},
\end{eqnarray*}
where $\theta^{(k)}:= \frac{\pi}{4}$ if $k \notin \mathcal{Q}$.
\end{lem}

The lemma is easily proved by taking advantage of the expression~(\ref{eq:schmidt}) for the Schmidt decomposition of a cluster state with one subsystem being qubit $k$ alone, and then calculating $\rho_k$ directly. The calculation is done by expressing the cluster state as the action of controlled-$\Z$ gates acting on the product state $\ket{+}^{\otimes n}$, and then using the fact that the $D^{(i)}$ and controlled-$\Z$ gates are mutually commuting. A consequence of this lemma is that the reduced density matrix of a given qubit is maximally mixed if and only if the qubit itself and at least one of its neighbors are untouched by \B-type operators. In general, qubits within \B-transformed cluster states are not maximally entangled with the rest of the state; in fact, they can be arbitrarily weakly entangled (with respect to the von Neumann entanglement entropy).
%A consequence of this lemma is that the reduced density matrix of a given qubit is maximally mixed if and only if the qubit itself and all of its neighbors are untouched by \B-type operators. 

Consider now a ring of an even number of qubits, with identical \B-type operators specified by $\theta^{(2k)}=\theta$ acting on the qubits with even labels. The significance of such a state will become clear in Example~\ref{ex:modstrategy2}. Two-point correlation functions can be calculated exactly for this state. 
%The algebra is tedious but straightforward, and makes heavy use of the fact that the operators $B^{(2k)\dag} B^{(2k)}$ commute with $\mathrm{CZ}$ gates. 
The result is that the nearest-neighbor correlation functions $C_{2k,2k\pm1}(\Z,\X) = \cos{2\theta}$ and the next-nearest-neighbor correlation functions $C_{2k,2k\pm2}(\Z,\Z) = C_{2k-1,2k+1}(\X,\X) = \cos^2(2\theta)$ are the only ones that are non-zero, while all the other two-point Pauli correlation functions are identically zero. Again, this differs from the perfect cluster state, where all correlation functions are zero.

Exact calculations on chains of up to 7 qubits with identical (but arbitrary) 
$\B$-type operators $\B_{2j}=D\R_z\left[\gamma^{(2j)}\right]$ acting on even 
qubits $2j$ reveal that the non-zero Schmidt coefficients corresponding to any 
bipartition of the chain into two contiguous halves are 
\set{\cos{\theta},\sin{\theta}}. The von Neumann entropy of entanglement is 
equal to the Shannon entropy of this list, and is generally less than one ebit.
The MPS representation bears out this observation. In the canonical form the 
site matrices for the boundary qubits are 
\begin{eqnarray}
A^{[1]}[0] &=& \frac{1}{\sqrt{2}}(\cos{\theta^{(2)}}\bra{0} +\sin{\theta^{(2)}}\bra{1}); \\
A^{[1]}[1] &=& \frac{1}{\sqrt{2}}(\cos{\theta^{(2)}}\bra{0} - \sin{\theta^{(2)}}\bra{1}); \\
A^{[n]}[0] &=&\ket{+}; \quad A^{[n]}[1] = -\ket{-},
\end{eqnarray}
those for the bulk even sites are
\begin{eqnarray}
A^{[2j]}[0] & = & \mtxtwotwo{0}{e^{-i\gamma^{(2j)}}}{0}{0};\\
A^{[2j]}[1] & = & \mtxtwotwo{0}{0}{e^{i\gamma^{(2j)}}}{0},
\end{eqnarray}
and those for the bulk odd sites are
\begin{eqnarray}
A^{[2j-1]}[0] & = & \mtxtwotwo{\cos{\theta^{(2j)}}}{\sin{\theta^{(2j)}}}{\cos{\theta^{(2j)}}}{\sin{\theta^{(2j)}}};\\
A^{[2j-1]}[1] & = & \mtxtwotwo{-\cos{\theta^{(2j)}}}{\sin{\theta^{(2j)}}}{\cos{\theta^{(2j)}}}{-\sin{\theta^{(2j)}}}.
\end{eqnarray}
It's very easy to verify that the channels induced by the matrices on the odd 
sites are not unital in the sense given in Ref.~\cite{Gross2010}, so a
$\B-\U-\B$ chain is not a quantum wire.

Such a 1D state would appear not to be capable of reliably processing a single 
qubit. This is true, but a simple modification of the geometry from one to two
dimensions yields a usable resource for random length computation. This will be
elaborated upon in Example~\ref{ex:modstrategy2}.

\section{Random Length Computation}
\label{sec:randomlength}
Neither Strategy I nor Strategy II discussed in the previous section directly offers a way to perform deterministic single-qubit rotations. For Strategy I, it is unclear how to compensate for a byproduct operator \rx{\eta} where $\eta \neq \pi$, as such a byproduct operator does not possess convenient commutation properties with the $\H$ and $\mathrm{R_z}$ operations. Similarly, for Strategy II, it is unclear whether some number of non-unitary teleported gates can be combined to form a desired unitary. 

Another perspective on the strategies is that a single measurement with outcome 1 teleports the gate $\hrz{\xi + \epsilon}$, where $\xi \in \field{C}$ is some angle associated with the always-on operation $\hrz{\xi}$ (in the terminology of Ref.~\cite{Gross2010}) and $\epsilon \in \field{C}$ is a possibly complex error associated with the byproduct. To correct this error in principle requires two additional measurement steps. The first measurement step should teleport the gate $\hrz{0} \equiv \H$, which would cancel the previously applied Hadamard gate; a possible $\X$ byproduct operator might result depending on the measurement outcome. On the second measurement step one would attempt to teleport $\hrz{-\epsilon}$ or $\hrz{\epsilon}$ depending on the previous measurement outcome, thus cancelling the original error $\epsilon$. 

This procedure is only possible if the measurement immediately after first incurring an error cannot itself generate any further error $\epsilon'$. One way to guarantee such a circumstance is to impose that every alternate $S_i$ operator is in fact unitary. Thus there must exist a class of states that are a strict subset of SLOCC-transformed cluster states, which constitute resources for random-length universal gate teleportation. Likewise, a subset of SLOCC-transformed cluster states in two dimensions must be universal resources for MBQC. The remainder of this section is devoted to various explicit examples. 

%\begin{exam}{\textbf{Deterministic single-qubit rotations: }$\mathbf{\N}-\mathbf{\U}-\mathbf{\N}$ \textbf{state.}}\\
\subsection{Deterministic single-qubit rotations: $\mathbf{\N}-\mathbf{\U}-\mathbf{\N}$ state}
\label{ex:exam1}

\begin{figure}[t,scale=0.5]
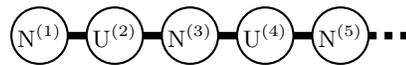

	\include{drawstrategyI}
	\caption{$\mathrm{N}-\mathrm{U}-\mathrm{N}$ state, a one-dimensional structure that can be used for deterministic random-length single-qubit rotations.}
	\label{fig:NUN}
\end{figure}
Consider a one-dimensional state of the form
\begin{equation}
\ket{R} = \N^{(1)}_1 \otimes \U^{(2)}_2 \otimes \N^{(3)}_3 \otimes \U^{(4)}_4 
\dots \otimes \N^{(n)}_n \cl{n},
\end{equation}
where the \set{N^{(i)}} are \N-type operators, and the \set{U^{(j)}} are local unitaries (c.f. Fig.~\ref{fig:NUN}). The goal is to teleport the single-qubit unitary
\begin{equation}
\U \brackets{\zeta, \eta, \xi} = \rx{\zeta} \rz{\eta} \rx{\xi}.
\end{equation}
The first step is to use Strategy I to attempt a teleportation of \H\rz{0}, 
analogously to the scheme with the perfect 1D cluster state. For the correct 
choice of basis the measurement outcome $m_1=0$ corresponds to success. One 
can then immediately measure qubit 2 in a basis that teleports 
$X^{m_2} \H \rz{\xi}$, and then use Strategy I to attempt the teleportation of 
$\H \rz{\brackets{-1}^{m_2}\eta}$ starting on qubit 3. 

If the first measurement outcome is instead $m_1=1$ then one instead teleports 
$\H\rz{\epsilon}$ with 
$\epsilon \in \field{R}$. This error must be immediately corrected, because the
next desired rotation is around an orthogonal axis. Happily, there is a local 
unitary $\U^{(2)}$ acting on the next qubit in the chain. The Hadamard operator
that effects the now-undesired transformation of the rotation axes can be 
eliminated by teleporting another one ($\H^2=\I$). This is accomplished by 
measuring the next qubit in the basis $\set{\U^{(2)}\ket{+}, \U^{(2)} \ket{-}}$. 
Labelling the measurement outcome $m_2$, the teleported gates are
\begin{equation}
\X^{m_2} \H \H \rz{\epsilon} \equiv \X^{m_2} \rz{\epsilon}.
\end{equation}
The measurement basis for qubit 3 is then chosen such that measurement outcome 
$m_3=0$ results in the gate $\H \rz{\brackets{-1}^{m_2+1} \epsilon}$ being 
teleported. In this case, the overall unitary becomes
\begin{eqnarray*}
\H \rz{\brackets{-1}^{m_2+1} \epsilon} X^{m_2} \rz{\epsilon} & = &  Z^{m_2} \H \rz{-\epsilon} \rz{\epsilon} \\
& = & Z^{m_2} \H.
\end{eqnarray*}
At this point one has successfully teleported a Hadamard gate and an 
unimportant Pauli byproduct. The next measurement on a qubit with an even 
label can teleport the desired $\H \rz{\xi}$ gate without error. One then 
attempts to teleport $\H \rz{\eta}$ by measuring qubit 5, using Strategy I, 
etc. 

The procedure corresponds to the following steps:

\begin{enumerate}

\item Measure qubit 1 with outcome $m_1$ in the basis 
\begin{equation}
\set{(\mathrm{N}^{(1) \dagger})^{-1} \rz{-\xi_1}\H \ket{0},\mathrm{N}^{(1)}\rz{-\xi_1}\H \ket{1}};
\label{eq:Nbasis}
\end{equation}

\item If $m_1 = 0$, then success;

\item If $m_1=1$ then one has effectively teleported the gate $\rx{\epsilon^{(1)}} \H\rz{\xi_1}$, where
\begin{equation}
\epsilon^{(1)} = \pm 2 \arctan{\frac{\cos{2\theta^{(1)}}\cos{\xi_1}}{1 \pm \cos{2\theta^{(1)}}\sin{\xi_1}}}+\pi.
\label{eq:epsilon}
\end{equation}
Note that $\epsilon^{(1)}=0$ when $\N^{(1)}=\U^{(1)}$ ($\theta^{(1)}=\pi/4$), as expected. Measure qubit 2 with outcome $m_2$ in the basis \set{\mathrm{U}^{(2)}\X\ket{0},\mathrm{U}^{(2)}\X\ket{1}};

%\item Measure qubit 3 with outcome $m_3$ in the basis $\set{(\mathrm{N}^{(3) \dagger})^{-1} \rz{(-1)^{m_2}\epsilon^{(1)}}\H \ket{0},	\mathrm{N}^{(3)} \rz{(-1)^{m_2}\epsilon^{(1)}}\H \ket{1}}$;

\item Measure qubit 3 with outcome $m_3$ in the basis $\set{(\mathrm{N}^{(3) \dagger})^{-1} \rz{\chi}\H \ket{0},	\mathrm{N}^{(3)} \rz{\chi}\H \ket{1}}$, 
where $\chi=(-1)^{m_2}\epsilon^{(1)}$;

\item Repeat steps 3 and 4 on successive qubits $2k$ and $2k+1$ until outcome 1 is achieved on an odd qubit, using $\mathrm{U}^{(2)} \rightarrow \mathrm{U}^{(2k)}$, $\mathrm{N}^{(3)} \rightarrow \mathrm{N}^{(2k+1)}$, $m_2 \rightarrow m_{2k}$, $\epsilon^{(1)} \rightarrow \epsilon^{(2k-1)}$.

\end{enumerate}
The key point of this example is that as for any measurement 
on an odd-numbered qubit that yields the `correct' outcome $m_i=0$, one will 
have succeeded in implementing part of the desired single-qubit rotation. 
Furthermore, any errors resulting from outcomes $m_i=1$ are correctible by 
making further measurements. This thus constitutes a repeat-until-success 
strategy, and gives rise to a quasi-deterministic random-length single-qubit 
rotation. The likely reason for this one-dimensional state to be capable of processing
a logical qubit is that the left and right parts of the state share an ebit of entanglement
with respect to any cut, as mentioned in Sec.~\ref{sec:ntransformed}.

%Some insight into this phenomenon can be gleaned by looking at the 
%entanglement properties of this state, as discussed in Sec.... [some 
%links to the previous (not yet written) discussion of the properties of 
%N-transformed clusters are needed here.]

%\end{exam}

\subsection{Deterministic Universal MBQC: 2D $\mathbf{\N-\U-\N}$ State}
\label{sec:2DNUN}

For universal MBQC, a two-dimensional resource state is required. The precise geometry of the two-dimensional state on which MBQC occurs is determined by the specific circuit to be implemented. Ideally one would start with a state defined on a convenient and simple geometry, and then `carve' the desired shape out by deleting certain qubits. For cluster-state MBQC, for example, one carves the required state out of a rectangular lattice by projectively measuring the unwanted qubits in the computational basis. The goal is to yield isolated one-dimensional wires, each of which represents a logical qubit, with links only existing between wires in places where an entangling gate between logical qubits is needed. 

\begin{figure}
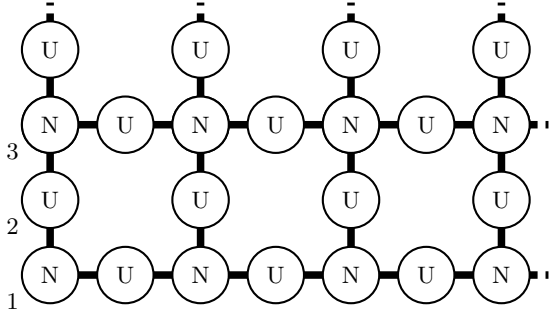

	\include{2DNUN}
	\caption{2D $\N-\U-\N$ state, universal for quasi-deterministic MBQC. The qubits labeled $1$, $2$ and $3$ can be used to implement an entangling gate, which involves measuring qubit 2 in the $\Y$ basis. Alternatively, if an entangling gate is not desired here, qubits $1$ and $3$ can be decoupled by measuring qubit $2$ in the $\Z$ basis. }
	\label{fig:2DNUN}
\end{figure}

Consider now a regular two-dimensional lattice composed of $\mathbf{\N-\U-\N}$ 
states, as depicted in Fig.~\ref{fig:2DNUN}. As in the usual cluster state,
logical qubits are processed by alternating horizontal wires composed of 
physical qubits, and entangling gates by vertical chains connecting them. 
Unlike the cluster case, however, the procedure for implementing single-qubit 
rotations with $\mathbf{\N-\U-\N}$ states is of random length, so it is 
impossible to decide in advance where the desired links between wires will 
occur. The computational cluster state then must be carved `on the fly.'

%The procedure for universal MBQC on two-dimensional $\mathbf{\N-\U-\N}$ states is as follows. 
Suppose that the quantum information encoding two logical qubits resides on 
(yet unmeasured) N-transformed physical qubits on two different wires. If an 
entangling gate between logical qubits is not desired at the next step, then 
the link between the wires can be first severed by measuring the intervening 
U-transformed chain qubit in the computational basis. An example of the method 
to decouple qubits 1 and 3 is shown in Fig.~\ref{fig:2DNUN}, where qubit 2 is
measured in the computational basis. This has the effect of teleporting a Z 
gate to each of the logical qubits if the measurement outcome is $m=1$. Other 
than taking into account the possible existence of these byproduct operators, 
the computation subsequently proceeds as in the one-dimensional case discussed 
above. 

The desired entangling gate is implemented as follows. At the time that an 
entangling gate is needed, the local part of the resource state looks like two 
1D $\N-\U-\N$ states, each coupled via $\mathrm{CZ}$ operations to an ancilla
initially in the state $\ket{+}$ and subsequently acted on by an arbitrary U.
In Fig.~\ref{fig:2DNUN}, the entangling link is represented by the vertical
$\N-\U-\N$ chain labeled by qubits 1, 2, and 3. The local part of the state is 
mathematically described as
\begin{equation}
\ket{R} = \N^{(1)}_1 \U^{(2)}_2 \N^{(3)}_3 \mathrm{CZ}_{1,2} \mathrm{CZ}_{2,3} \ket{\mathrm{c}+\mathrm{t}}_{123},
\end{equation}
where the states \ket{\mathrm{c}} and \ket{\mathrm{t}} could be thought of as
control and target states respectively for some entangling gate. 
Now, qubits 1 and 3 are measured in the usual Strategy I 
basis~(\ref{eq:Nbasis}) with $\xi^{(i)}=0$, while qubit 2 is measured in the 
eigenbasis of the Pauli operator \Y, suitably rotated by $\U^{(2)}$. This 
procedure teleports the state initially situated on qubits 1 and 3 through an 
entangling gate 
\begin{eqnarray}
\nonumber G_{1,3} &=& \rx{\mu^{(1)}}^{m_1}_1 \X_1^{m_1+m_2} \H_1 \nonumber \\
&\times& \rx{\mu^{(3)}}^{m_3}_3 \X_3^{m_2+m_3} \H_3 M_{1,3}
\end{eqnarray}
to qubits 4 and 5, with 
\begin{equation}
M_{1,3} = \ketbra{00}{00}_{1,3} + i \ketbra{01}{01}_{1,3} + i \ketbra{10}{10}_{1,3} + \ketbra{11}{11}_{1,3}.
\end{equation}
Here, the $\set{\mu^{(i)}}$ are the standard Strategy I byproduct angles, 
Eq.~(\ref{eq:byproductangle}) or (\ref{eq:epsilon}). This entangling operation is related to $\mathrm{CZ}$ via 
\begin{equation}
\mathrm{CZ}_{1,3} \equiv \X_1 \X_3 \rz{\pi/2}_1 \rz{\pi/2}_3 M_{1,3} \X_1 \X_3,
\end{equation}
and so $G_{i,j}$ together with single-qubit operators forms a universal set of gates. 
%\end{exam}

%\begin{exam}{\textbf{Probabilistic single-qubit rotations: }$\mathbf{\B}-\mathbf{\U}-\mathbf{\B}$ \textbf{state.}}\\
\subsection{Probabilistic single-qubit rotations: $\mathbf{\B}-\mathbf{\U}-\mathbf{\B}$ state}
\begin{figure}[t,scale=0.5]
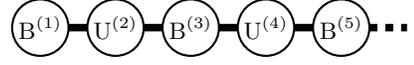

	\include{drawstrategyII}
	\caption{$\mathrm{B}-\mathrm{U}-\mathrm{B}$ state, a one-dimensional structure that can be used for probabilistic random-length single-qubit rotations.}
	\label{fig:BUB}
\end{figure}
Next consider a one-dimensional state of the form
\begin{equation}
\label{eq:rdef} \ket{R} = \B^{(1)}_1 \otimes \U^{(2)}_2 \otimes \B^{(3)}_3 \otimes \U^{(4)}_4 \dots \otimes \B^{(2n+1)}_{2n+1} \cl{2n+1},
\end{equation}
where the \set{\B^{(2i+1)}} are \B-type operators, and the \set{\U^{(2i)}} are 
once again local unitaries (c.f.\ Fig.~\ref{fig:BUB}). This structure ensures 
that none of the bonds present in the structure is perfect; no particle has 
maximal entropy of entanglement with the rest of the state. This fact is a consequence of Lemma~\ref{lem:blocal}; there is no qubit unaffected by \B-type operators whose neighborhood contains any unaffected qubits.

All single-qubit 
measurements for the odd-numbered qubits now correspond to Strategy II, in 
which the byproduct operator is always $\X$ if it occurs. All even-numbered 
qubits are measured in the \set{\U^{(2i)}\ket{+},\U^{(2i)}\ket{-}} basis; as in 
the previous example, the only purpose of these measurements is to enable the 
removal of undesired contributions to the rotation angles. The main difference 
from the previous example is that a non-unitary gate of the form 
$\H \rz{\xi_{2i+1}}$ is teleported, where $\xi_{2i+1} \in \field{C}$. The 
present goal is therefore to compensate for the imaginary part of the rotation 
angle. 

As discussed in the previous section and Eq.~(\ref{eq:strategy2table}), the 
imaginary part of $\xi_{2i+1}$ is 
entirely determined by the ratio of the singular values of $\B^{(2i+1)}$, and
can be defined as $\epsilon=i\ln\left(\cot\theta^{(2i+1)}\right)$.
Consider momentarily the special case where the \set{\B^{(2i+1)}} all have the 
same singular values. The gate teleported by a measurement of 
the B-transformed qubit $2i+1$ will then be proportional to 
\rz{\epsilon(-1)^{m_{2i}+m_{2i-2}+\ldots}}, ignoring all rotations about real 
angles which are entirely determined by the choice of measurement basis. In
short, the sign of the imaginary angle depends on the outcomes of the previous 
measurements on even-numbered U-transformed qubits. 

The imaginary component 
therefore undergoes a random walk of step-length $|\epsilon|$.
In particular, the walker takes its first step to the right when the first measurement 
outcome of an even-numbered qubit is 0, and to the left if it is 1. Subsequently, a measurement outcome of 0 on an even-numbered qubit causes the walker to take another step in the same direction as the previous step, while outcome 1 makes the walker take a step in the opposite direction. 
%When $\theta\neq\pi/4$, the probabilities are such that on any step the walker tends to prefer to proceed in the same direction as before rather than turning around.

The two possible measurement outcomes with odd qubits are always equally 
likely, but the probabilities with even qubits depend on the singular values 
of the $\B$-type operators from the (odd) neighboring qubits, and generally speaking the walker is more likely to stray further from the origin than to step back towards it. For example, consider measuring the first two qubits of $\ket{R}$ in Eq.~(\ref{eq:rdef}) in the $\set{\ket{+},\ket{-}}$ basis. It can easily be shown, using the Schmidt decomposition~(\ref{eq:schmidt}) and the expression~(\ref{eq:bchar}), that the probabilities of the outcome $\ket{\pm}$ on qubit 1 are equal, and that those on qubit 2 are

\begin{equation}
p_{\pm,2} = \frac{1}{2} \brackets{1 \pm \cos{2\theta^{(1)}}\cos{2\theta^{(3)}}}.
\end{equation}

Here, the random walk effectively begins at position $\mathrm{ln} \brackets{\tan{\theta^{(1)}}}$ and moves to $\mathrm{ln} \brackets{\tan{\theta^{(1)}}} \pm \mathrm{ln} \brackets{\tan{\theta^{(3)}}}$. For a situation to arise where the walker moves closer to the origin with probability greater than $1/2$, one of the two pairs of conditions
\begin{eqnarray}
\label{eq:walkcond11} \abs{\mathrm{ln}\brackets{\tan{\theta^{(1)}}} \pm \mathrm{ln}\brackets{\tan{\theta^{(3)}}}}&>&2\abs{\mathrm{ln}\brackets{\tan{\theta^{(1)}}}}; \\
\label{eq:walkcond12} \pm \cos{2\theta^{(1)}}\cos{2\theta^{(3)}}&>&0
\end{eqnarray}
must be simultaneously satisfied, for either sign. If the $\set{\theta^{(i)}}$ are chosen uniformly at random, then the probability of this happening is only about $0.315$. If this measurement procedure is continued down the chain, with qubits 3-5 relabelled 1-3 after the first two measurements and so on, the current value of $\theta^{(1)}$ tends to drift away from $\pi/4$ towards either $0$ or $\pi/2$, and the range of values of $\theta^{(3)}$ for which the walker is likely to turn around and walk towards the origin progressively shrinks. Furthermore, if $\set{\theta^{(1)}}$ and $\set{\theta^{(3)}}$ are equal at any time, the walker is guaranteed to be more likely to continue in one direction than to turn around.
%The probability that the walker will take a step back in the direction of the origin is always lower than the probability it will travel further away, and 
%these probabilities become increasingly skewed with distance from the origin. 
%The walker is reasonably likely (although with probability less than 1/2) to return to the origin after the measurement of qubit 2, but if it does not then it becomes less and less likely that the walker will ever return after this. 

This procedure constitutes a probabilistic method for implementing a 
single-qubit rotation. Unfortunately if the walker strays too far from the
origin, it becomes effectively impossible to recover and the attempted gate 
teleportation fails. The entire computation must then be repeated. If the 
singular values of the $\B^{2i+1}$ are chosen such that the imaginary 
components of the teleported angles are all integer multiples of each other, 
then the behavior of the random walk is even more deleterious. A judicious 
two-dimensional arrangement of $\B - \U - \B$ chains avoids this catastrophe, 
as discussed in the next example.

%\begin{table*}
%	\centering
%	\small
%	\begin{tabular*}{0.95\textwidth}{c | p{0.35\textwidth} p{0.5\textwidth}}
%	\toprule 
%		Step & Action & Measurement basis \\
%	\hline
%	1 & Measure qubit 1, outcome $m_1$. & \set{\rz{\xi-\alpha^{(1)}+e^{i\beta^{(1)}}\frac{\pi}{2}}\rx{\frac{\pi}{2}}\ket{0},\rz{\xi-\alpha^{(1)}+e^{i\beta^{(1)}}\frac{\pi}{2}}\rx{\frac{\pi}{2}}\ket{1}} \\
%	2 & Measure qubit 2, outcome $m_2$. & \set{\mathrm{U}^{(2)}\X\ket{0},\mathrm{U^{(2)}}\X\ket{1}}\\
%	3 & Measure qubit 3, outcome $m_3$. & \set{\rz{-\alpha^{(3)}+e^{i\beta^{(3)}}\frac{\pi}{2}}\rx{\frac{\pi}{2}}\ket{0},\rz{\xi-\alpha^{(3)}+e^{i\beta^{(3)}}\frac{\pi}{2}}\rx{\frac{\pi}{2}}\ket{1}}\\
%	4 & If $m_2$ = 1, success. & -\\
%	5 & Repeat steps 3 and 4 on successive qubits $2k$ and $2k+1$ until difference between number of 1 and 0 outcomes on even-numbered qubits is 1. & $\mathrm{U}^{(2)} \rightarrow \mathrm{U}^{(2k)}$,$m_2 \rightarrow m_{2k}$, $\alpha^{(3)} \rightarrow \alpha^{(2k+1)}$,$\beta^{(3)}\rightarrow \beta^{(2k+1)}$.  \\
%	\hline
%	\end{tabular*}
%	\caption{Procedure for probabilistically teleporting a single-qubit unitary $\H \rz{\xi_1}$ with a $\mathrm{B}-\mathrm{U}-\mathrm{B}$ state after a random (possibly infinite) number of steps.}
%\end{table*}
%\end{exam}

%\begin{exam}{\textbf{Universal MBQC: Modified} $\mathbf{\B}-\mathbf{\U}-\mathbf{\B}$ \textbf{state in 3D.}}
\subsection{Universal MBQC: Percolated 2D Cluster State from 3D $\mathbf{\B}-\mathbf{\U}-\mathbf{\B}$ state}
\label{ex:modstrategy2}
In the previous example using Strategy II, a possibly infinite number of steps 
may be required to teleport an arbitrary single-qubit unitary. But quitting the
protocol results in catastrophic failure: because the computational wire is
effectively broken, the entire gate teleportation must be attempted from the 
beginning. A solution to these problems is to employ the 3D extension to the 
previous resource, corresponding to a cluster state transformed by alternating 
$\mathrm{B}$-type operators and unitaries. This corresponds to a lattice with 
two interpenetrating cubic sublattices, a \B-lattice and a \U-lattice.

%One 2D plane of this 3D resource is depicted in Fig.~\ref{fig:2D}. The planes
%immediately above and below the page have the same structure, but with the 
%$\B$ and $\U$ labels interchanged; two planes away the pattern returns. 
%Initially, $\Z$-basis measurements within a given
%plane are used to carve out a structure in which each \B-transformed qubit of 
%a $\B - \U -\B$ horizontal computational wire is attached to a long vertical
%chain of $\B - \U -\B$ states. Measurements are made on the chain qubits, 
%starting at the \B-transformed qubit on the computational wire and continuing 
%in the vertical direction until success (defined below) is achieved. The goal 
%is to probabilistically produce perfect entanglement along the computational
%wire, thereby effectively eliminating the $\B$ operators in the horizontal
%direction.

\begin{figure}
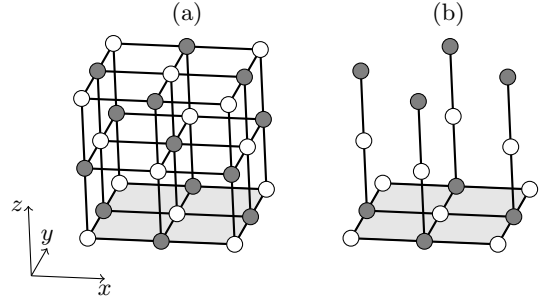

	\begin{minipage}{0.9\columnwidth}
	\begin{center}
		\include{draw2Dresource4}
	\end{center}
	\end{minipage}
\caption{3D cluster state with $\mathrm{B}$-type operators and unitaries 
acting on alternate qubits (a) before carving and (b) after carving. Grey qubits are acted upon by $\B$-type operators and white qubits by local unitaries. $\Z$-basis measurements are made in the $z$ 
direction in (a) to disentangle those vertical chains originating from a white qubit in the $x-y$ plane. A measurement protocol along the remaining 
vertical chains in (b) produces a percolated 2D cluster in the $x-y$ plane.}
	\label{fig:2D}
\end{figure}

An example of this 3D resource, a cube with side length 3, is depicted in Fig.~\ref{fig:2D}. 
Initially, $\Z$-basis measurements in the $z$-direction (as labeled in Fig.~\ref{fig:2D})
are used to carve out a structure in which each \B-transformed qubit in the $x-y$ plane,
shaded grey, is attached to a long vertical $\B - \U -\B$ chain. Measurements are made on the chain qubits, 
starting at qubit above the \B-transformed qubit on the computational wire and continuing 
in the vertical direction until success (defined below) is achieved. The goal 
is to probabilistically produce perfect entanglement in the $x-y$ plane, thereby effectively eliminating the $\B$ operators in the horizontal
direction. The result is a 2D cluster state in this plane with 
missing entanglement bonds in random locations. As long as the mean density of 
broken links exceeds the percolation threshold for a two-dimensional square 
lattice, the resource is universal for MBQC~\cite{Kieling2007b}.

%Failure in this protocol again corresponds to a broken computational wire, so
%there doesn't seem to be any improvement over the previous example. But it 
%should be kept in mind that the same sequence is carried out in every plane of 
%the 3D resource. 

%Again for simplicity assume that all $\B$-type operators are identical, with 
%the ratio of their singular values denoted by $\lambda=\cot(\theta)$. Also, 
%the unitary operators acting on the other qubits are ignored; they can simply 
%be incorporated into the measurement bases used for the measurements of those 
%qubits. 
Consider the first vertical chain from 
the left in Fig.~\ref{fig:2D}. Recall that one can interpret $\B$ as a 
$z$-rotation by an imaginary angle $\pm i\ln\lambda$, as shown in 
Eq.~(\ref{eq:BtypeRz}). For simplicity, we assume the unitary operators acting on even-numbered qubits are all equal to the identity; were they not, they could be compensated by a suitable rotation of the measurement basis for. The portion of the state corresponding to the first 
four qubits of the vertical chain (with the qubit that intersects the 
horizontal chain labeled 1), is then (ignoring normalization)
\begin{eqnarray*}
\ket{T}&=&\B_1 \B_3 \mathrm{CZ}_{1,2}\mathrm{CZ}_{2,3} \mathrm{CZ}_{3,4} \ket{++++}_{1234} \\
& = & \rz{i \mathrm{ln}\lambda}_1 \rz{i \mathrm{ln}\lambda}_3
\ket{\mbox{Cl}_4}_{1234},
\end{eqnarray*}
ignoring normalization factors as usual. First, qubits 2 and 3 are measured in the \set{\ket{+},\ket{-}} basis (of course, the measurement basis for qubit 2 would need to be rotated if a local unitary $\U^{(2)}$ were acting). These are commuting measurements, since they are on different qubits and not 
adaptive. The effect is to teleport the state $\rz{i \mathrm{ln} \lambda}\ket{+}$ from qubit 3 through $\X^{m_2}\H\X^{m_1}\H \equiv \X^{m_2}\Z^{m_1}$ to qubit 1, yielding the new state
\begin{eqnarray*}
\ket{T'}&=&\rz{i \mathrm{ln}\lambda}_1 \mathrm{CZ}_{1,4} \X^{m_2}_1 \Z^{m_1}_1 \rz{i \mathrm{ln}\lambda}_1 \ket{++}_{14} \\
& = & \mathrm{CZ}_{1,4} \rz{i ((-1)^{m_2}+1)\mathrm{ln}\lambda)}_1 \ket{++}_{14}
\end{eqnarray*}
up to overall local unitaries on the final state. 

If $m_2 = 1$, then the 
imaginary part of the rotation angle is completely canceled. Qubit 4 can then
be measured in the computational basis (again, suitably rotated if necessary) to disentangle the rest of the vertical 
chain from the horizontal chain. The result is a perfect cluster along the 
first three qubits in the horizontal direction, and the $\B$-type operator is
effectively deleted. 

If $m_2=0$, then the situation is similar to the original. There is still a 
$\B$-type operator present in the horizontal direction, now corresponding to a 
z-rotation about an angle with imaginary part 
$2 \mathrm{ln} \lambda = \mathrm{ln} \lambda^2$. In other words, the new 
effective $\B$-type operator in the horizontal chain has a ratio of singular 
values that is the square of the original one. In order to remove the effect 
of the $\B$-type operator, the chain qubits must be measured sequentially 
until the total number of steps towards the origin exceeds by 1 the total number of steps away. 

The probabilities $p_0^{(k)}$ and $p_1^{(k)}$ of the outcomes 0 and 1 on an even qubit in the vertical 
chain,where $k > 0$ is the present position of the walker on the real number line, are given by
\begin{eqnarray}
\label{eq:zeroprob} p_0^{(k)} & = & \frac{1+\lambda^{2k+2}}{1+\lambda^2+\lambda^{2k}+\lambda^{2k+2}} \sim O(1);\\
\label{eq:oneprob} p_1^{(k)} & = &\frac{\lambda^2+\lambda^{2k}}{1+\lambda^2+\lambda^{2k}+\lambda^{2k+2}} \sim O(\lambda^2).
\end{eqnarray}
The total probability $p_n$ that the effect of the $\B$ will be undone within 
$2n$ measurements is the sum of the probabilities of all of the possible trajectories of the walker on $n$ or fewer steps with initial position 1, final position 0 and all intermediate positions strictly positive.

\begin{figure}
\includegraphics[scale=1]{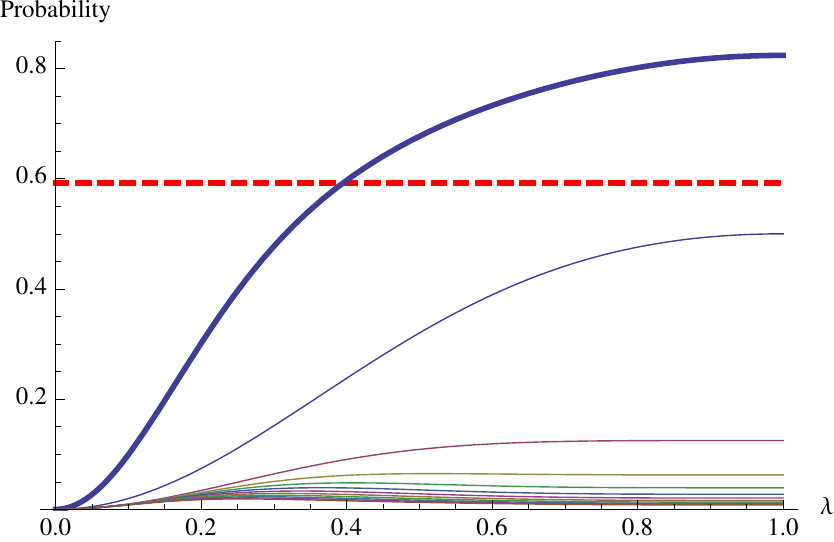}
\caption{(Color online) Probability of success of the procedure for deleting a $\B$-type operator in the plane via a random walk in the third dimension, as a function of the ratio $\lambda$ of the singular values (thick blue, color online). The probabilities $p_k$ of deleting $\B$ with exactly $k$ even-qubit measurements are also shown for $k$ from 1 to 10 (thin, decreasing with increasing $k$), and the thick blue line is the sum of these. The red dashed line is the percolation threshold.}
\label{fig:comparison}
\end{figure}

The probability $p_{10}$ of undoing the B operator after 10 attempts (20 measurements)
is shown in Fig.~\ref{fig:comparison} as a function of the ratio of singular 
values $\lambda$. Calculation of the exact probability $p_{\infty}$ is computationally intractable, for two reasons. First, the number of valid trajectories for the walker grows exponentially in the number of steps allowed. Second, the probability of any particular trajectory depends on the full history of the walker, not just the number of steps. Of course, $p_{\infty}$ must approach 1 as $\lambda$ approaches unity (the limit
that B becomes a unitary matrix). In this case, the walk reduces to the simple 1D random walk, which is known to sample the origin frequently.

If after some predetermined number of measurements along a vertical chain one
has not yet succeeded in undoing B, the qubit at the root of the chain (i.e.\ 
in the computational wire) can be measured in the computational basis and 
thereby deleted. The result is a broken link in the 2D cluster state. 
The important result shown in Fig.~\ref{fig:comparison} is that there is a 
critical value of $\lambda$, called $\lambda_c$, above which the probability of successfully 
undoing the $\B$ rises above the (bond) percolation threshold for a 2D square 
lattice (approximately 0.593). For this walk, the critical value obeys $\lambda_c \lesssim 0.379$. The upper bound for $\lambda_c$ is read off the thick blue curve from Fig.~\ref{fig:comparison}. Thus, this 
procedure probabilistically yields a universal resource for MBQC provided 
that $\lambda$ is sufficiently large. We note that a similar example was considered in~\cite{Mora2010}, where the resource was a 2D cluster state with identical $\B$-type operators acting everywhere and the percolation proceeded via two-element POVMs that either removed the $\B$ or deleted the qubit. There, the critical value of $\lambda$ was found to be $0.649$.

%The lower bound is estimated as follows. First, note that the probability of the walker returning to the origin in \textit{exactly} $n$ steps is a monotonically increasing function of $\lambda$ in the region $\lambda \in \brackets{0,1}$ (see inset of Fig.~\ref{fig:comparison}). It then follows that the difference between the true probability and $p_{10}$ is also monotonically increasing. Since $p_{\infty} \rightarrow 1$ as $\lambda \rightarrow 1$, and $\lim_{\lambda \rightarrow 1} p_{10} = 0.832$, the difference between $p_{10}$ and $p_{\infty}$ must be less than $0.168$ for all $\lambda$. Thus, the lower bound corresponds to the value of $\lambda$ for which $p_{10} = 0.593-0.168$.

%\end{exam}

\begin{section}{Discussion and Conclusions}
\label{sec:discussion}
Motivated by a desire to identify new resource states for measurement-based quantum computing, we have performed a (non-exhaustive) search of the equivalence class of $n$-qubit cluster states on a rectangular lattice, under the action of $\gl^{\otimes n}$. In particular, our aim was to identify which states within this class could be used as resources for MBQC, by designing explicit protocols for teleporting single-qubit gates and two-qubit entangling gates, driven by adaptive local projective measurements. We identified a class of one-dimensional states, the so-called $\mathrm{N}-\mathrm{U}-\mathrm{N}$ states, that are deterministically universal for single-qubit rotations, although with a random number of measurements needed to teleport the desired rotation. We also identified a probabilistically universal resource for single-qubit rotations: the so-called $\mathrm{B}-\mathrm{U}-\mathrm{B}$ states. We then described a three-dimensional extension of $\mathrm{B}-\mathrm{U}-\mathrm{B}$ states that can yield a universal resource for deterministic MBQC beyond a percolation threshold, and a 2D $\N-\U-\N$ state that is also universal for deterministic but random-length MBQC.

Several interesting open issues arise as the result of this work. First, it is
not clear what (if any) relationship exists between the states uncovered in 
this work and other known resource states. For example, the probabilistic 
nature of the protocol with $\B-\U-\B$ states has features in common with that 
of other resources for MBQC, such as photonic cluster states prepared via 
probabilistic entangling gates or with unreliable 
sources~\cite{Gross2006,Kieling2007,Kieling2007a,Kieling2010}. 
Likewise, the quasi-deterministic $\mathrm{N}-\mathrm{U}-\mathrm{N}$ states
share various characteristics with the AKLT-inspired resources of 
Refs.~\cite{Gross2007a,Brennen2008,Bartlett2010,Mora2010}, in particular the
exponential spin correlations and the repeat-until-success measurement-based 
strategies. One distinction is that only one alternating sublattice of the $\N-\U-\N$ states exhibits non-zero correlation functions, whereas in the AKLT chain, every qubit is correlated with every other. Presumably a true identification of an AKLT-type resource with a $\N$-transformed cluster state will require $\N$-type operators to be present on every qubit, a case we have not handled here.

Also, it will be important to better understand the 
relationships of these states with the universal quantum wires of 
Ref.~\cite{Gross2010}. In that work, certain reasonable physical assumptions 
were imposed on 1D wires at the outset, for example: the possibility of producing a wire via a translationally invariant nearest-neighbour global entangling operation, the asymptotic sharing of an ebit of entanglement between the left and right halves of the chain, etc. In our work, it is not clear if a translationally invariant scheme exists for producing $\N-\U-\N$ or $\B-\U-\B$ chains. Evidence from exact calculations on small chains and the explicit description of the states within the MPS representation reveals that although the left and right halves of $\N-\U-\N$ chains share an ebit, the halves of the $\B-\U-\B$ chains do not. This seems reasonable, as the $\N-\U-\N$ chain is quasideterministically universal for single-qubit rotations, while the $\B-\U-\B$ chain is only probabilistically so. 

%It is not clear from our work if 
%$\mathrm{N}-\mathrm{U}-\mathrm{N}$ or $\mathrm{B}-\mathrm{U}-\mathrm{B}$ 
%states necessarily satisfy these properties, as we did not impose them. 

Second, it is conceivable that all states that have been hitherto identified as
universal resources for MBQC are in fact SLOCC-equivalent to the family of 
cluster states. There is some evidence to support this conjecture. For 
example, the results of Ref.~\cite{Chen2010} show that many seemingly diverse 
resource states can be reduced to cluster states via local strategies. 
Similarly, the proof of the universality of the 2D AKLT state on a honeycomb 
lattice proceeds via local reduction to a random graph state, which can in 
turn be reduced to a percolated cluster state~\cite{Wei2011}. This reduction 
is successful despite the fact that the initial resource is defined on qutrits 
rather than qubits, and on a non-rectangular lattice. An even more intriguing
possibility (though we believe it to be unlikely) is that all possible states 
for universal MBQC fall within the SLOCC-equivalence class of the cluster 
states. At the very least, the relative size of this class decreases 
exponentially with the total number of physical qubits, as 
expected~\cite{Gross2009,Bremner2009}.

Third, while we have shown that a certain subset of the orbit of the cluster 
states under SLOCC are useful resources, either probabilistically or 
quasi-deterministically, it is not clear if the remaining SLOCC-transformed 
cluster states are also universal resources for MBQC. Generically, single-qubit
measurements on these states teleport gates with byproduct operators that are 
rotations about the X axis by complex angles. One possibility is that there is
a measurement protocol that can accommodate all possible byproducts that we
simply haven't found. Perhaps there is another sense, besides 
quasi-deterministic or probabilistic, in which these states can be said to be 
useful for MBQC. Another possibility is that there is some map from the full 
orbit to the particular subset of states considered in the work. Alternatively, 
the states in the full orbit may not be useful resources, though we have not
attempted to prove this. 
 
Finally, it would be useful if the resources presented in this work could 
be realized as the ground states of a physical Hamiltonian. A recent no-go 
theorem~\cite{Chen2011} shows this cannot be the case for frustration-free 
Hamiltonians on qubits. The question remains open for frustrated Hamiltonians, 
for instance the AKLT Hamiltonian in the Haldane 
phase~\cite{Affleck1987,Affleck1988,Bartlett2010}. Another possibility is that
the ground states of physical Hamiltonians could be locally reduced to the 
resources we have found. Further research is needed to answer these and
related questions.
\end{section}

\begin{section}{Acknowledgements}
The authors are grateful to Jens Eisert, David Gross, and Gilad Gour for 
stimulating discussions. This research was supported by Alberta Innovates - Technology Futures and the Natural Sciences and Engineering Research Council of Canada (NSERC). 
\end{section}

\bibliography{wires9.bbl}
%\bibliographystyle{apsrev}
%\bibliography{PhD3.bib}
%\bibliography{wires9}
\end{document}

%% file: drawstrategyI.tex
										\begin{tikzpicture}[thick,scale=0.5]
											%\draw[fill=green!20,rounded corners=2mm] (0,0) rectangle (11,-2);
											\foreach \j in {1,...,5}
											{
												\path (2*\j-1,-1) coordinate (qc\j);
												\path (2*\j+1,-1) coordinate (qc2\j);
												\draw[fill=white] (qc\j) circle (0.75);
											}
											\foreach \k in {1,2,3,4}
											{
												\path (qc\k) ++ (0.75,0) coordinate (temp\k);
												\path (qc2\k) ++ (-0.75,0) coordinate (temp2\k);	
												\draw[line width=1mm] (temp\k) -- (temp2\k);
											}
											\path (qc5) ++ (0.75,0) coordinate (temp5) ++ (1,0) coordinate (temp25);
											\draw[line width=1mm,dashed] (temp5) -- (temp25);
											\node (lab1) at (qc1) {$\mathrm{N^{(1)}}$}; 
											\node (lab2) at (qc2) {$\mathrm{U^{(2)}}$}; 
											\node (lab3) at (qc3) {$\mathrm{N^{(3)}}$}; 											
											\node (lab4) at (qc4) {$\mathrm{U^{(4)}}$}; 
											\node (lab5) at (qc5) {$\mathrm{N^{(5)}}$};
										\end{tikzpicture}

%% file: 2DNUN.tex
			\begin{tikzpicture}[thick,scale=0.5]
				%\useasboundingbox (0,0) rectangle (15,9);
				%\draw[fill=green!20,rounded corners=2mm] (0,0) rectangle (15,9);
				\foreach \j in {1,...,7}
				{
					\foreach \k in {1,3} 
					{
						\path (2*\j-1,2*\k-1) coordinate (qc\j\k);
						\path (2*\j+1,2*\k-1) coordinate (qc2\j\k);
						\path (2*\j-1,2*\k+1) coordinate (qcv2\j\k);
						\path (qc\j\k) ++ (0,-0.75) coordinate (below\j\k);
						\draw[fill=white] (qc\j\k) circle (0.75);
					}
				}
				\foreach \j in {1,3,5,7}
				{
					\foreach \k in {2,...,4} 
					{
						\path (2*\j-1,2*\k-1) coordinate (qc\j\k);
						\path (2*\j+1,2*\k-1) coordinate (qc2\j\k);
						\path (2*\j-1,2*\k+1) coordinate (qcv2\j\k);
						\path (qc\j\k) ++ (0,-0.75) coordinate (below\j\k);
						\draw[fill=white] (qc\j\k) circle (0.75);
					}
				}
				\foreach \j in {1,3,5,7}
				{
					\foreach \k in {1,...,3}
					{
						\path (qc\j\k) ++ (0,0.75) coordinate (tempv\j\k);
						\path (qcv2\j\k) ++ (0,-0.75) coordinate (tempv2\j\k);
						\draw[line width=1mm] (tempv\j\k) -- (tempv2\j\k);
					}					
				}
				\foreach \j in {1,3,5,7}
				{
					\foreach \k in {4,...,4}
					{
						\path (qc\j\k) ++ (0,0.75) coordinate (tempv\j\k);
						\path (qcv2\j\k) ++ (0,-0.75) coordinate (tempv2\j\k);
						\draw[line width=1mm,dashed] (tempv\j\k) -- (tempv2\j\k);
					}					
				}
				\foreach \j in {1,...,6}
				{
					\foreach \k in {1,3}
					{
						\path (qc\j\k) ++ (0.75,0) coordinate (temp\j\k);
						\path (qc2\j\k) ++ (-0.75,0) coordinate (temp2\j\k);
						\draw[line width=1mm] (temp\j\k) -- (temp2\j\k);
					}
				}
				\foreach \j in {7,...,7}
				{
					\foreach \k in {1,3}
					{
						\path (qc\j\k) ++ (0.75,0) coordinate (temp\j\k);
						\path (qc2\j\k) ++ (-0.75,0) coordinate (temp2\j\k);
						\draw[line width=1mm,dashed] (temp\j\k) -- (temp2\j\k);
					}
				}
				\foreach \j in {1,3,5,7}
				{
					\foreach \k in {1,3}
					{
						\node (lab\j\k) at (qc\j\k) {$\mathrm{N}$};
					}
				}
				\foreach \j in {2,4,6}
				{
					\foreach \k in {1,3}
					{
						\node (lab\j\k) at (qc\j\k) {$\mathrm{U}$};
					}
				}
				\foreach \j in {1,3,5,7}
				{
					\foreach \k in {2,4}
					{
						\node (lab\j\k) at (qc\j\k) {$\mathrm{U}$};
					}
				}
				\foreach \j in {1,...,3}
				{
					\path(qc1\j) ++ (-1,-0.7) coordinate (labpos\j);
					\node(lab\j) at (labpos\j) {$\j$};
				}
			\end{tikzpicture}

%% file: drawstrategyII.tex
										\begin{tikzpicture}[thick,scale=0.5]
											%\draw[fill=green!20,rounded corners=2mm] (0,0) rectangle (11,-2);
											\foreach \j in {1,...,5}
											{
												\path (2*\j-1,-1) coordinate (qc\j);
												\path (2*\j+1,-1) coordinate (qc2\j);
												\draw[fill=white] (qc\j) circle (0.75);
											}
											\foreach \k in {1,2,3,4}
											{
												\path (qc\k) ++ (0.75,0) coordinate (temp\k);
												\path (qc2\k) ++ (-0.75,0) coordinate (temp2\k);	
												\draw[line width=1mm] (temp\k) -- (temp2\k);
											}
											\path (qc5) ++ (0.75,0) coordinate (temp5) ++ (1,0) coordinate (temp25);
											\draw[line width=1mm,dashed] (temp5) -- (temp25);
											\node (lab1) at (qc1) {$\mathrm{B^{(1)}}$}; 
											\node (lab2) at (qc2) {$\mathrm{U^{(2)}}$}; 
											\node (lab3) at (qc3) {$\mathrm{B^{(3)}}$}; 											
											\node (lab4) at (qc4) {$\mathrm{U^{(4)}}$}; 
											\node (lab5) at (qc5) {$\mathrm{B^{(5)}}$};
										\end{tikzpicture}

%% file: draw2Dresource4.tex
% Sketch output, version 0.3 (build 2d, Thu Jun 9 14:59:18 2011)
% Output language: PGF/TikZ,LaTeX
\begin{tikzpicture}[line join=round]
\filldraw[fill=gray!20](0,0)--(-.423,-.732)--(1.531,-.813)--(1.953,-.081)--cycle;
\filldraw[fill=gray!20](3.5,0)--(3.077,-.732)--(5.031,-.813)--(5.453,-.081)--cycle;
\draw[thick](-.041,.93)--(-.081,1.859);
\draw[thick](-.041,.93)--(-.252,.564);
\draw[thick](-.041,.93)--(.936,.889);
\draw[thick](-.081,1.859)--(-.292,1.493);
\draw[thick](-.081,1.859)--(.895,1.819);
\draw[thick](-.211,-.366)--(-.252,.564);
\draw[thick](-.211,-.366)--(-.423,-.732);
\draw[thick](-.211,-.366)--(.765,-.407);
\draw[thick](-.252,.564)--(-.292,1.493);
\draw[thick](-.252,.564)--(-.463,.198);
\draw[thick](-.252,.564)--(.725,.523);
\draw[thick](-.292,1.493)--(-.504,1.127);
\draw[thick](-.292,1.493)--(.684,1.453);
\draw[thick](-.423,-.732)--(-.463,.198);
\draw[thick](-.423,-.732)--(.554,-.773);
\draw[thick](-.463,.198)--(-.504,1.127);
\draw[thick](-.463,.198)--(.513,.157);
\draw[thick](-.504,1.127)--(.473,1.087);
\draw[thick](.473,1.087)--(1.449,1.046);
\draw[thick](.513,.157)--(.473,1.087);
\draw[thick](.513,.157)--(1.49,.117);
\draw[thick](.554,-.773)--(.513,.157);
\draw[thick](.554,-.773)--(1.531,-.813);
\draw[thick](.684,1.453)--(.473,1.087);
\draw[thick](.684,1.453)--(1.661,1.412);
\draw[thick](.725,.523)--(.513,.157);
\draw[thick](.725,.523)--(.684,1.453);
\draw[thick](.725,.523)--(1.701,.483);
\draw[thick](.765,-.407)--(.554,-.773);
\draw[thick](.765,-.407)--(.725,.523);
\draw[thick](.765,-.407)--(1.742,-.447);
\draw[thick](.895,1.819)--(.684,1.453);
\draw[thick](.895,1.819)--(1.872,1.778);
\draw[thick](.936,.889)--(.725,.523);
\draw[thick](.936,.889)--(.895,1.819);
\draw[thick](.936,.889)--(1.913,.849);
\draw[thick](.977,-.041)--(.765,-.407);
\draw[thick](.977,-.041)--(.936,.889);
\draw[thick](.977,-.041)--(1.953,-.081);
\draw[thick](0,0)--(-.041,.93);
\draw[thick](0,0)--(-.211,-.366);
\draw[thick](0,0)--(.977,-.041);
\draw[thick](1.49,.117)--(1.449,1.046);
\draw[thick](1.531,-.813)--(1.49,.117);
\draw[thick](1.661,1.412)--(1.449,1.046);
\draw[thick](1.701,.483)--(1.49,.117);
\draw[thick](1.701,.483)--(1.661,1.412);
\draw[thick](1.742,-.447)--(1.531,-.813);
\draw[thick](1.742,-.447)--(1.701,.483);
\draw[thick](1.872,1.778)--(1.661,1.412);
\draw[thick](1.913,.849)--(1.701,.483);
\draw[thick](1.913,.849)--(1.872,1.778);
\draw[thick](1.953,-.081)--(1.742,-.447);
\draw[thick](1.953,-.081)--(1.913,.849);
\draw[thick](3.077,-.732)--(4.054,-.773);
\draw[thick](3.248,.564)--(3.208,1.493);
\draw[thick](3.289,-.366)--(3.077,-.732);
\draw[thick](3.289,-.366)--(3.248,.564);
\draw[thick](3.289,-.366)--(4.265,-.407);
\draw[thick](3.5,0)--(3.289,-.366);
\draw[thick](3.5,0)--(4.477,-.041);
\draw[thick](4.013,.157)--(3.973,1.087);
\draw[thick](4.054,-.773)--(4.013,.157);
\draw[thick](4.054,-.773)--(5.031,-.813);
\draw[thick](4.265,-.407)--(4.054,-.773);
\draw[thick](4.265,-.407)--(5.242,-.447);
\draw[thick](4.436,.889)--(4.395,1.819);
\draw[thick](4.477,-.041)--(4.265,-.407);
\draw[thick](4.477,-.041)--(4.436,.889);
\draw[thick](4.477,-.041)--(5.453,-.081);
\draw[thick](5.201,.483)--(5.161,1.412);
\draw[thick](5.242,-.447)--(5.031,-.813);
\draw[thick](5.242,-.447)--(5.201,.483);
\draw[thick](5.453,-.081)--(5.242,-.447);
\filldraw[fill=gray](-.041,.93) circle (3pt);
\filldraw[fill=gray](-.211,-.366) circle (3pt);
\filldraw[fill=gray](-.292,1.493) circle (3pt);
\filldraw[fill=gray](-.463,.198) circle (3pt);
\filldraw[fill=gray](.473,1.087) circle (3pt);
\filldraw[fill=gray](.554,-.773) circle (3pt);
\filldraw[fill=gray](.725,.523) circle (3pt);
\filldraw[fill=gray](.895,1.819) circle (3pt);
\filldraw[fill=gray](.977,-.041) circle (3pt);
\filldraw[fill=gray](1.49,.117) circle (3pt);
\filldraw[fill=gray](1.661,1.412) circle (3pt);
\filldraw[fill=gray](1.742,-.447) circle (3pt);
\filldraw[fill=gray](1.913,.849) circle (3pt);
\filldraw[fill=gray](3.208,1.493) circle (3pt);
\filldraw[fill=gray](3.289,-.366) circle (3pt);
\filldraw[fill=gray](3.973,1.087) circle (3pt);
\filldraw[fill=gray](4.054,-.773) circle (3pt);
\filldraw[fill=gray](4.395,1.819) circle (3pt);
\filldraw[fill=gray](4.477,-.041) circle (3pt);
\filldraw[fill=gray](5.161,1.412) circle (3pt);
\filldraw[fill=gray](5.242,-.447) circle (3pt);
\filldraw[fill=white](-.081,1.859) circle (3pt);
\filldraw[fill=white](-.252,.564) circle (3pt);
\filldraw[fill=white](-.423,-.732) circle (3pt);
\filldraw[fill=white](-.504,1.127) circle (3pt);
\filldraw[fill=white](.513,.157) circle (3pt);
\filldraw[fill=white](.684,1.453) circle (3pt);
\filldraw[fill=white](.765,-.407) circle (3pt);
\filldraw[fill=white](.936,.889) circle (3pt);
\filldraw[fill=white](0,0) circle (3pt);
\filldraw[fill=white](1.449,1.046) circle (3pt);
\filldraw[fill=white](1.531,-.813) circle (3pt);
\filldraw[fill=white](1.701,.483) circle (3pt);
\filldraw[fill=white](1.872,1.778) circle (3pt);
\filldraw[fill=white](1.953,-.081) circle (3pt);
\filldraw[fill=white](3.077,-.732) circle (3pt);
\filldraw[fill=white](3.248,.564) circle (3pt);
\filldraw[fill=white](3.5,0) circle (3pt);
\filldraw[fill=white](4.013,.157) circle (3pt);
\filldraw[fill=white](4.265,-.407) circle (3pt);
\filldraw[fill=white](4.436,.889) circle (3pt);
\filldraw[fill=white](5.031,-.813) circle (3pt);
\filldraw[fill=white](5.201,.483) circle (3pt);
\filldraw[fill=white](5.453,-.081) circle (3pt);
\path (-.845*0.5,-1.464*0.5) ++ (-0.75,-0.5) coordinate (origin);
\path (1.108*0.5,-1.545*0.5) ++ (-0.75,-0.5) coordinate (x);
\path (x) ++ (0,-0.15) coordinate (belowx);
\path (-.423*0.5,-.732*0.5) ++ (-0.75,-0.5) coordinate (y);
\path (y) ++ (0,0.15) coordinate (righty);
\path (-.926*0.5,.395*0.5) ++ (-0.75,-0.5) coordinate (z);
\path (z) ++ (-0.15,0) coordinate (leftz);
\draw[->] (origin) -- (x); 
\draw[->] (origin) -- (y); 
\draw[->] (origin) -- (z); 
\node (labx) at (belowx) {$x$};
\node (laby) at (righty) {$y$};
\node (labz) at (leftz) {$z$};
\node (laba) at (0.9,2.25) {(a)};
\node (labb) at (4.375,2.25) {(b)};
\end{tikzpicture}% End sketch output